\documentclass[12pt,final,href]{iopart}
\usepackage[dvips]{color}
\usepackage{graphicx}
\usepackage{citesort}

\makeatletter

\begin{document}

\def\GeV{{\rm GeV}}
\def\lesssim{\mathrel{\hbox{\rlap{\hbox{\lower4pt\hbox{$\sim$}}}\hbox{$<$}}}}
\def\lsim{\lesssim}
\def\gsim{\mathrel{\rlap{\lower4pt\hbox{\hskip1pt$\sim$}} \raise1pt\hbox{$>$}}}
\def\der{d}
\def\TMC{{\rm TMC}}
\def\comment#1{\textcolor{red}{#1}}
\def\comment#1{\textcolor{red}{\large #1}}

%%%%%%%%%%%%%%%%%%%%%%%%%%%%%%%%%%%%%%%%%%%%%%%%%%%%%%%%%%%%
% DEFINE XI/ETA NOTATION:
\newcommand{\xieta}{\ensuremath{\xi}}
\newcommand{\xietabar}{\ensuremath{\bar\xi}}
%%%%%%%%%%%%%%%%%%%%%%%%%%%%%%%%%%%%%%%%%%%%%%%%%%%%%%%%%%%%

%\tableofcontents{}

\newpage{}

%%%%%%%%%%%%%%%%%%%%%%%%%%%%%%%%%%%%%%%%%%%%%%%%%%%%%%%%%%%%%%%%%%%
%%%%%%%%%%%%%%%%%%%%%%%%%%%%%%%%%%%%%%%%%%%%%%%%%%%%%%%%%%%%%%%%%%%
%%%%%%%%%%%%%%%%%%%%%%%%%%%%%%%%%%%%%%%%%%%%%%%%%%%%%%%%%%%%%%%%%%%
%\input{0cover}
\newpage{}

%%%%%%%%%%%%%%%%%%%%%%%%%%%%%%%%%%%%%%%%%%%%%%%

%\null \hfill  arXiv:YYMM.NNNNvV \\
%\null \hfill arch-ive/yymmnnn \\
%\null \hfill arXiv:YYMM.NNNN \\
\null \hfill arXiv:0709.1775 [hep-ph]\\
\null \hfill JLAB-THY-07-718 
%\null \hfill hep-ph/yymmnnn
%\null \hfill arch-ive/yymmnnn

\title[A Review of Target Mass Corrections]{A Review of Target Mass 
Corrections\footnote{J. Phys. G: Nucl. Part. Phys. 35 (2008) 053101}}

\author{
Ingo~Schienbein\rlap,${}^{a,b}$ 
Voica~A.~Radescu\rlap,${}^{c}$
G.P.~Zeller\rlap,${}^{d}$            
M.~Eric~Christy\rlap,${}^{e}$         
C.E.~Keppel\rlap,${}^{e,f}$          
Kevin~S.~McFarland\rlap,${}^{g}$           
W.~Melnitchouk\rlap,${}^{f}$             
Fredrick~I.~Olness\rlap,${}^{b}$\footnote{Corresponding 
  author e-mail: olness@smu.edu} 
Mary~Hall~Reno\rlap,${}^{h}$\footnote{Corresponding 
  author e-mail: mary-hall-reno@uiowa.edu}                 
Fernando~Steffens\rlap,${}^{i}$        
Ji-Young Yu\rlap,${}^{b}$        
}

\address{
\mbox{${}^a$Laboratoire de Physique Subatomique \& Cosmologie, F-38026 Grenoble, France }\\ 
\mbox{${}^b$Southern Methodist University, Dept. of Physics, Dallas, TX 75275, USA   }\\ 
\mbox{${}^c$Deutsches Elektronen Synchrotron, Notkestrasse 85, D-22603 Hamburg, Germany   }\\ 
\mbox{${}^d$Los Alamos National Laboratory, Los Alamos, NM 87545, USA}\\ 
\mbox{${}^e$Hampton University,  Dept. of Physics, Hampton, VA 23668, USA  }\\ 
\mbox{${}^f$Jefferson Lab, Newport News, VA 23606, USA}\\ 
\mbox{${}^g$University of Rochester, Dept. of Physics \& Astronomy, Rochester, NY 14627-0171, USA   }\\ 
\mbox{${}^h$University of Iowa, Dept. of Physics and Astronomy, Iowa City, IA 52242, USA  }\\ 
\mbox{${}^i$Mackenzie Presbiteriana Universidade, 01302-907, Sao Paulo, SP, Brazil   }\\ 
}

\date{\today}

\begin{abstract}
With recent advances in the precision of 
inclusive lepton--nuclear scattering 
experiments,
it has become apparent that comparable improvements are needed in the
accuracy of the theoretical analysis tools.  In particular, when extracting parton
distribution functions in the large-$x$ region, it is crucial to correct the data
for effects associated with the nonzero mass of the target.  We present here a
comprehensive review of these target mass corrections (TMC) to structure
functions data, summarizing the relevant formulas for TMCs in electromagnetic
and weak processes.  We include a full analysis of both hadronic and partonic
masses, and trace how these effects appear in the operator product expansion
and the factorized parton model formalism, as well as their limitations when
applied to data in the $x \to 1$ limit.  We evaluate the numerical effects of TMCs
on various structure functions, and compare fits to data with and without these
corrections.
\end{abstract}

% \pacs{possible choices **************
% \\
% 13.60.Hb 	Total and inclusive cross sections (including deep-inelastic processes)\\
% 12.39.St 	Factorization\\
% 12.38.-t 	Quantum chromodynamics\\
% 11.15.Bt 	General properties of perturbation theory\\
%  }

\maketitle

%%%%%%%%%%%%%%%%%%%%%%%%%%%%%%%%%%%%%%%%%%%%%%%
%\newpage

%%%%%%%%%%%%%%%%%%%%%%%%%%%%%%%%%%%%%%%%%%%%%%%%%%%%%%%%%%%%%%%%%%%
%%%%%%%%%%%%%%%%%%%%%%%%%%%%%%%%%%%%%%%%%%%%%%%%%%%%%%%%%%%%%%%%%%%
%%%%%%%%%%%%%%%%%%%%%%%%%%%%%%%%%%%%%%%%%%%%%%%%%%%%%%%%%%%%%%%%%%%
\tableofcontents{}
\newpage
\markboth{A Review of Target Mass Corrections}{A Review of Target Mass Corrections}

%%%%%%%%%%%%%%%%%%%%%%%%%%%%%%%%%%%%%%%%%%%%%%%%%%%%%%%%%%%%%%%%%%%
%%%%%%%%%%%%%%%%%%%%%%%%%%%%%%%%%%%%%%%%%%%%%%%%%%%%%%%%%%%%%%%%%%%
%%%%%%%%%%%%%%%%%%%%%%%%%%%%%%%%%%%%%%%%%%%%%%%%%%%%%%%%%%%%%%%%%%%
%\newpage{}
%\input{1intro-3}

\section{Introduction}
\label{sec:intro}

The scattering of electrons on hadronic targets has historically
played an essential role in our understanding of the proton as
a composite particle made up of partons: quarks and gluons
\cite{Taylor:1991ew,Kendall:1991np,Friedman:1991nq,Devenish:2004pb}.
Presently, data from electron and neutrino scattering at large momentum
transfers, that is, deeply inelastic scattering (DIS), are used to determine
the parton distribution functions (PDFs) which characterize the substructure
of hadrons \cite{Gluck:1998xa,Lai:1999wy,Kretzer:2003it,Alekhin:2006zm,%
Blumlein:2006be,Martin:2007bv}. 
At lower energies, the resonant components of hadronic structure,
and the duality between hadronic and partonic descriptions of interactions,
continue to be explored \cite{Bloom:1970xb,Melnitchouk:2005zr}.

As the precision of the recent lepton--hadron scattering data has improved,
it is vital for the theoretical analysis to keep pace. For example, the
calculation of the Wilson coefficients has progressed to encompass
next-to-leading order (NLO) quantum chromodynamics (QCD) and beyond
\cite{Altarelli:1979ub,Furmanski:1981cw,Gluck:1987uk,%
Gluck:1996ve,Vermaseren:2005qc,Moch:2004xu,Vogt:2004mw,Moch:2004pa,%
Zijlstra:1992kj,Zijlstra:1992qd}.
It is important, therefore, to consider all sources of corrections which may
contribute at a comparable magnitude, such as electroweak radiative
corrections
\cite{Diener:2005me,Martin:2004dh}, 
quark mass effects \cite{Barnett:1976ak,Witten:1975bh,Gottschalk:1980rv,%
Collins:1986mp,Aivazis:1993pi,Buza:1996wv,Chuvakin:1999nx,%
Kretzer:1998ju}, 
and target mass corrections \cite{Nachtmann:1973mr,%
Georgi:1976ve,Barbieri:1976rd,Barbieri:1976bj,%
Ellis:1982wd,Ellis:1982cd,DeRujula:1976ih,DeRujula:1976tz}.
In this review,
we will focus on the problem of target mass corrections (TMCs), which formally
are subleading $1/Q^2$ corrections to leading twist structure functions, where
$Q^2$ is the squared four-momentum transfer to the hadron.

Understanding TMCs is important for several reasons.
Their effects are most pronounced at large $x$ and moderate $Q^2$,
which coincides with the region where parton distribution functions (PDFs)
are not very well determined.  A reliable extraction of PDFs here therefore
demands an accurate description of the TMCs.  Furthermore, a reliable
interpretation of data on multiparton correlations at low momentum transfer
depends on the proper accounting of TMCs.
While target mass corrections have a long history, implementing these
has not been entirely straightforward, as there exist a number
of conventions, prescriptions, and potential scheme choices which
can lead to differences in the final numerical results.

The target mass corrections to electroweak structure functions were
first determined by Georgi and Politzer in 1976 \cite{Georgi:1976ve}
within the operator product expansion (OPE) at the leading order of QCD.
In the same year Barbieri et al.
\cite{Barbieri:1976rd,Barbieri:1976bj} rederived the mass corrections to
scaling in DIS, including effects arising from non-zero quark masses.
These same corrections were later derived from a parton model approach
by Ellis, Furmanski and Petronzio \cite{Ellis:1982wd,Ellis:1982cd}.
Beyond leading order, the NLO QCD corrections to the target mass corrected
structure functions
were derived by De R\'ujula, Georgi and Politzer \cite{DeRujula:1976ih}.
Recently, Kretzer and Reno \cite{Kretzer:2002fr,Kretzer:2003iu} reevaluated
the TMCs for charged current (CC) and neutral current (NC) $\mu$- and $\tau$-
neutrino DIS, including NLO QCD corrections.

There is a number of theoretical ingredients necessary to derive the TMCs
to hadronic structure functions in lepton--hadron scattering in the context of
the OPE, in order to relate them to the quark-parton model (QPM).
The OPE method makes use of basic fundamental symmetries to relate
the cross section to a reduced matrix element; as such, the OPE takes
the hadron mass fully into account.
In order to relate the reduced matrix elements to quantities which can be
computed in the QPM, one then inherits the associated limitations.
For example, the QPM describes the interaction as involving the scattering
from a single, free parton (the leading twist contribution, see Sec.\ \ref{sec:OPE}).
Multi-parton correlations, formally higher twist, are discarded.
Additionally,
the QPM also imposes collinear kinematics on the parton involved in the
interaction. The potential for the parton momentum to have a non-zero component 
transverse to the hadron momentum vector is neglected. This effect can
omit mass contributions and introduce ambiguities if not addressed carefully.

When using the QPM approach, the target mass corrections are taken
into account in the following places: 

\begin{itemize}

\item in relating the parton fraction of the hadron's light-cone momentum 
(called the Nachtmann variable) to the Bjorken scaling variable $x=Q^2/2M\nu$,
where $M$ is the hadron mass and $\nu$ the energy transfer;

\item in the mixing between the partonic and hadronic structure functions;

\item in a collinear expansion, where the TMCs appear from the $p_{T}$ effects\ .

\end{itemize}

In the limit of negligible target mass relative to $Q^2$, the Nachtmann variable
reduces to Bjorken-$x$. The QPM approach including TMCs has further
limitations in the limit as $x \to 1$. A ``threshold problem'' arises when trying
to enforce that the structure functions vanish in kinematically forbidden regions.

In the next section, we review structure functions and the OPE approach. 
We show the OPE results for the
structure functions. Section~3 discusses the relation of the OPE to the
parton model. We illustrate the formalism with the example of neutrino
charged current scattering in Section~4. Section~5 describes
the $x\rightarrow 1$ problem and recent attempts to resolve this
theoretical issue. Finally, in Section~6, we make some numerical
comparisons using structure functions with and without target mass 
corrections. 
Our conclusions are summarized in Section~7.
Three appendices detail the notation used in this review,
the inclusion of charm mass corrections in neutrino charged current
scattering, and a comparison of notation and results with Refs.\ 
\cite{Georgi:1976ve,Barbieri:1976rd,DeRujula:1976tz} and 
\cite{Kretzer:2003iu}.

%%%%%%%%%%%%%%%%%%%%%%%%%%%%%%%%%%%%%%%%%%%%%%%%%%%%%%%%%%%%%%%%%%%
%%%%%%%%%%%%%%%%%%%%%%%%%%%%%%%%%%%%%%%%%%%%%%%%%%%%%%%%%%%%%%%%%%%
%%%%%%%%%%%%%%%%%%%%%%%%%%%%%%%%%%%%%%%%%%%%%%%%%%%%%%%%%%%%%%%%%%%
%\input{2ope}
%\input{2ope-rev-4}

\section{Structure Functions and the Operator Product Expansion}
\label{sec:OPE}

In this section we present an overview of the formalism for deep inelastic
scattering (DIS).  Using the framework of the operator product expansion
(OPE), we outline the derivation of the leading electromagnetic and weak
structure functions at finite $Q^2$, including corrections from the non-zero
target nucleon mass.  We summarize these in a set of so-called ``master
equations''.  Finally, we present an alternative but equivalent formulation
of target mass correction (TMC) effects through the Nachtmann moments,
and contrast this with the formulation in terms of Cornwall--Norton moments.

\subsection{Overview of Structure Functions and the OPE}

%---------------------------------------------
\begin{figure}[!t]
\begin{center}
\includegraphics[scale=0.80,angle=0]{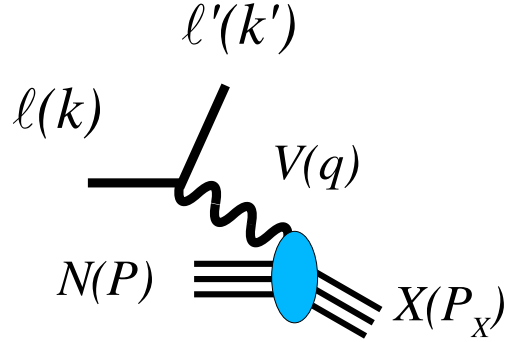} 
\caption{The basic deep inelastic lepton--nucleon scattering
	process, $\ell(k)+N(P) \to \ell'(k')+X(P_{X})$.}
\label{fig:DIS} 
\end{center}
\end{figure}
%---------------------------------------------

The basic lepton--nucleon inelastic scattering process,
$\ell(k) + N(P) \to \ell'(k') + X(P_X)$, is shown schematically
in Fig. \ref{fig:DIS}, where $k (k')$ is the initial (final) lepton
four-momentum, $P$ is the target nucleon momentum,
and $P_X$ is the momentum of the final hadronic state $X$.
We define $q=k-k'$ to be the four-momentum transferred
from the lepton to the nucleon, with $Q^2 \equiv -q^2$.
The energies of the initial and final leptons are denoted
by $E$ and $E'$, respectively.
Our notation reserves $M$ for the target nucleon mass,
$P^2=M^2$, and the invariant mass squared of the final
hadronic state is given by
$P_X^2 = W^2 = (P + q)^2 = M^2 + 2 P \cdot q - Q^2$
(see also \ref{app:kin}).

For electromagnetic or weak neutral current (NC) scattering,
the vector boson ($V$)--nucleon subprocess is
$V(q) + N(P) \to X(P_X)$, where $V=\gamma ,\ Z^0$.
The related charged current (CC) process, which
is important in neutrino--hadron scattering,
$\nu(k) + N(p) \to \ell(k') + X(P_X)$, where $V=W^\pm$,
will be discussed in Sec.\ \ref{sec:OPEpm}, where we discuss the
correspondence with the parton model.

In addition to the virtuality of the exchanged boson, $Q^2$,
inelastic scattering is also characterized by the Bjorken
scaling variable $x$, where
\begin{equation}
x = \frac{Q^2}{2P\cdot q}\ .
\end{equation}
In the massless target and quark limits (or equivalently in the
$Q^2 \to \infty$ limit), $x$ is equivalent to the light-cone momentum
fraction of the target carried by the interacting parton.
In the target rest frame, the Bjorken variable can be written
$x = Q^2/2M\nu$, where $\nu=E - E'$  
is the energy transferred to the hadronic system, and we define the
inelasticity of the process by $y=\nu/E$. 
For convenience, we also introduce the variable $r$ to denote a
frequently appearing combination of factors:
\begin{equation}
r = \sqrt{1+\frac{4x^{2}M^{2}}{Q^{2}}}\equiv\sqrt{1+\frac{Q^{2}}{\nu^{2}}} .
\end{equation}

%---------------------------------------------
\begin{figure}[!t]
\begin{center}
\includegraphics[scale=0.50,angle=270]{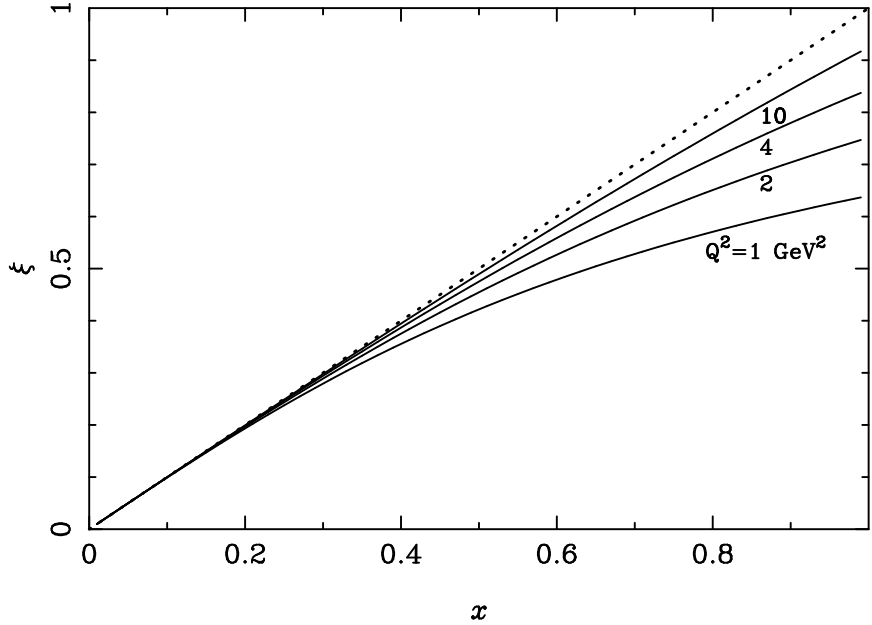} 
\caption{The Nachtmann variable $\xi$ as a function of the Bjorken
	scaling variable $x$, for $Q^2=1$, 2, 4 and 10~GeV$^2$.
For reference, a dotted line is shown for the limiting case 
$\xi=x$.
}
\label{fig:xi} 
\end{center}
\end{figure}
%---------------------------------------------

At finite $Q^2$, the effects of the target and quark masses modify the
identification of the Bjorken $x$ variable with the light-cone momentum
fraction.  For massless quarks, the parton light-cone fraction 
is given by the Nachtmann variable $\xieta$ \cite{Nachtmann:1973mr},
\begin{equation}
\xieta = \frac{2x}{1+\sqrt{1+4x^2M^2/Q^2}}\ .
\label{eq:xi}
\end{equation}
At large values of $Q^2$, $\xi \sim x$.
As Fig. \ref{fig:xi} shows, however, for $Q^2$ less than a few times
the target mass of $\sim 1$~GeV, $\xi$ can deviate significantly
from $x$, especially at large $x$ values.
The Nachtmann variable appears naturally in the OPE, as we
outline below. The full details of the notation, including parton
masses, appear in \ref{app:kin}.

We can write any generic 
inclusive lepton--nuclear scattering 
 cross section as a combination of a
hadronic tensor $W_{\mu\nu}$ and a leptonic tensor $L^{\mu\nu}$:
\[
{d\sigma}\sim\ W_{\mu\nu}\ L^{\mu\nu}\ ,
\]
where the hadronic tensor is given in terms of a product of hadronic
currents,\footnote{In this work, we will focus on the unpolarized results. For
TMC effects on the polarized structure function see
Refs.~\cite{Blumlein:1998an,Blumlein:1998nv,Blumlein:2006ia}, and
references therein.
}
\begin{eqnarray}
W_{\mu\nu}&\equiv&\frac{1}{2\pi}\int d^{4}z\ e^{iq\cdot z}\
	\langle N|[J_{\mu}(z),J_{\nu}(0)] |N\rangle\ \nonumber \\
&=& -g_{\mu \nu}W_1 + {p_\mu p_\nu\over M^2}W_2
	-i\epsilon_{\mu\nu\rho\sigma}
	{p^{\rho}q^\sigma\over M^2}W_3 \nonumber \\
&  &+ {q_\mu q_\nu \over M^2}W_4+{p_\mu q_\nu+p_\nu q_\mu\over M^2} W_5\ .
\end{eqnarray}
The structure functions $W_i$ depend on $x$ and $Q^2$, as well as the
target mass $M$.  The hadronic tensor can be related to the discontinuity
of the virtual forward Compton scattering amplitude $T_{\mu\nu}$ via 
\begin{equation}
\label{eq:disc}
W_{\mu\nu}=\frac{1}{\pi}\ {\rm disc}\, T_{\mu\nu}\ .
\end{equation}

The OPE allows one to expand the hadronic matrix element in the
forward scattering amplitude in a complete set of local operators
\cite{Miramontes:1988fz}:
\begin{eqnarray}
T_{\mu\nu}&\equiv& i\int d^{4}z\ e^{iq\cdot z}\
	\langle N | T[J_{\mu}(z)\, J_{\nu}(0)] | N \rangle \nonumber \\
& = & \sum_{i,\tau,n} c_{\tau,\mu\nu}^{i,\mu_1 \cdots \mu_n}(q)
	\langle N | O^{i,\tau}_{\mu_{1} \cdots \mu_{n}} | N \rangle \ ,
\end{eqnarray}
where the coefficient functions $c_{\tau,\mu\nu}^{i,\mu_1 \cdots \mu_n}(q)$
represent the hard scattering of the boson from the parton.
Here $\tau$ denotes the twist of the operator $O$, defined to be the
mass dimension minus the spin of the operator, and $i$ represents
different operators with the same twist.

In the approximation of keeping only twist-2 operators,\footnote{For 
a NLO calculation including an analysis of higher twist
contributions see, for example,
Refs.~\cite{Kataev:1999bp,Kataev:2001kk,Kataev:2005hv}
} 
the forward
scattering amplitude is explicitly:
\begin{eqnarray} \label{eq:tmunu}
T^{\mu\nu}
&&= \sum_{k=1}^\infty\Bigl( -g^{\mu\nu} q_{\mu_1}q_{\mu_2} C_1^{2k}
	+ g^\mu_{\mu_1}g^\nu_{\mu_2} Q^2 C_2^{2k} - i\epsilon ^{\mu\nu\alpha\beta}
	g_{\alpha\mu_1}q_\beta q_{\mu_2}C_3^{2k} \nonumber \\
&&
\phantom{= \sum_{k=1}^\infty\Bigl(}
+{q^\mu q^\nu\over Q^2}q_{\mu_1}q_{\mu_2} C_4^{2k}
	+( g^\mu_{\mu_1}q^\nu q_{\mu_2} + g^\nu_{\mu_1}q^\mu q_{\mu_2} )C_5^{2k}
	\Bigr) q_{\mu_3} \cdots q_{\mu_{2k}}
\nonumber\\
& & 
\phantom{=} \times
{2^{2k}\over Q^{4k}} A_{2k} \Pi^{\mu_1 \cdots \mu_{2k}}\ , 
%\nonumber\\
%& &
\end{eqnarray}
where $A_{2k}$ is the reduced matrix element of the twist-2 operator
of spin $2k$, and $C_i^{2k}$ is the Wilson coefficient calculated using
perturbative QCD.
In Eq.~(\ref{eq:tmunu}),
\begin{equation}
\Pi^{\mu_{1} \cdot \mu_{2k}}=\sum_{j=0}^{k}(-1)^{j}\ \frac{(2k-j)!}{2^{j}(2k)!}\ 
{{\underbrace{\{ g...g\}}\atop {\scriptstyle j\ g^{\mu_{n}\mu_{m}}{}'s}}}\quad
{{\underbrace{\{ p...p\}}\atop {\scriptstyle (2k-2j)\ p^{\mu_{n}}{}'s}}}(p^{2})^{j}\ ,
\label{eq:pi}
\end{equation}
where $\{ g...g\}\ \{ p...p\}$ abbreviates a sum over $(2k)!/[2^{j}j!(2k-2j)!]$
permutations of the indices.

The strategy in the calculation, as outlined by Georgi and Politzer \cite{Georgi:1976ve},
is to evaluate Eq.~(\ref{eq:tmunu}), picking off each of the coefficients in the expansion,
\begin{equation}\label{eq:t1etc}
T_{\mu\nu}= -g_{\mu \nu}T_1 + {p_\mu p_\nu\over M^2}T_2
-i\epsilon_{\mu\nu\rho\sigma}
{p^{\rho}q^\sigma\over M^2}T_3
+{q_\mu q_\nu \over M^2}T_4+{p_\mu q_\nu+p_\nu q_\mu\over M^2} T_5\ .
\end{equation}
Finally, the structure functions $W_i$ can be obtained from the imaginary parts
of the amplitudes $T_i$ using Eq.~(\ref{eq:disc}).
In modern notation, the structure functions are denoted by $F_i$ rather than $W_i$.
To emphasize the inclusion of target mass corrections in the structure functions,
we label $F_i \rightarrow F_i^{\TMC}$, and note the correspondence
\begin{eqnarray}
&& \biggl\{F_1,\ F_2,\ F_3,
\ F_4,\ F_5 \biggr\}^{\TMC}\nonumber\\
& & \quad = \left\{W_1, 
 \frac{Q^2}{2 x M^2}W_2 ,
 \frac{Q^2}{x M^2}W_3 ,
 \frac{Q^2}{2 M^2} W_4 ,
 \frac{Q^2}{2 x M^2} W_5
\right\}\ . 
\label{eq:WtoF}
\end{eqnarray}
In the following discussion, we focus on $F_1^{\TMC},\ F_2^{\TMC}$ and $F_3^{\TMC}$. 
The remaining two structure functions enter into the differential cross section
suppressed by the lepton mass squared divided by $ME$ \cite{Kretzer:2003iu},
and can for most purposes be neglected.

Using Eqs.~(\ref{eq:disc}) and (\ref{eq:tmunu}), one can explicitly relate the
Cornwall--Norton moments $M_i^n(Q^2)$ of the $F_i^{\TMC}$ structure functions
to sums of reduced matrix elements \cite{Georgi:1976ve,Solovtsov:1999in}.  
For the $F_2^{\TMC}$
structure function, for example, one has:
\begin{eqnarray}
\label{eq:f2moment}
M_2^n(Q^2) &=& \int_0^1 dx\, x^{n-2} F_2^{\TMC}(x,Q^2) \nonumber \\ 
&=& \sum _{j=0}^{\infty}
	\Biggl( \frac{M^2}{Q^2}\Biggr)^j\frac{(n+j)!}{j! (n-2)!}
	\frac{C_2^{n+2j}A_{n+2j}}{(n+2j)(n+2j-1)}\ .
\end{eqnarray}
Defining $F_i^{(0)}$ to be the massless nucleon limits of the structure
functions $F_i^{\rm TMC}$, we can relate the reduced matrix elements in
Eq.~(\ref{eq:f2moment}) to the Cornwall--Norton moments of $F_i^{(0)}$:
\begin{eqnarray}
C_2^{n+2j} A_{n+2j}
&\equiv& \int_0^1 dy\, y^{n+2j-2} F_2^{(0)}(y)\ ,
\label{eq:cnF2}	\\
C_i^{n+2j} A_{n+2j}
&\equiv& \int_0^1 dy\, y^{n+2j-1} F_i^{(0)}(y)\ ,\ \ \ \  i=1,3 \ .
\label{eq:cnF13}
\end{eqnarray}
In other words, the functions $F_i^{\rm TMC}$ contain target mass effects,
whereas the $F_i^{(0)}$ do not. 

The Cornwall--Norton moments 
can be inverted to yield a set of target mass corrected structure
functions derived from the operator product expansion:
\begin{eqnarray}
F_1^{\TMC}(x,Q^2)&=& \frac{x}{\xi r} F_1^{(0)}(\xi) - \frac{M^2 x^2}{Q^2}
\frac{\partial}{\partial x}\Biggl\{ \frac{1+r}{r} g_2(\xi ) \Biggr\}\ ,
\label{eq:partf1}
\\
F_2^{\TMC}(x,Q^2)&=&  x^2
\frac{\partial^2}{\partial x^2}\Biggl\{ \frac{(1+r)^2}{4 r} g_2(\xi ) \Biggr\}\ ,
\label{eq:partf2}\\
\label{eq:partf3}
F_3^{\TMC}(x,Q^2) & = & -x\frac{\partial}{\partial x}
\Biggl( \frac{1+r}{2r}
h_3( \xi ) \Biggr)\ .
\end{eqnarray}
The $F_2^{\TMC}$-expression comes from Eq.~(4.17) of Ref.~\cite{Georgi:1976ve},
and $F_1^{\TMC}$ and $F_3^{\TMC}$ follow in the same manner.
The functions $g_2$ and $h_3$ along with two more auxiliary functions
$h_{1,2}$, which will be needed below, are given by \cite{Kretzer:2003iu}:
\begin{eqnarray}
h_{1}(\xieta,Q^{2}) & = & \int_{\xieta}^{1}du\ \frac{2F_{1}^{(0)}(u,Q^{2})}{u}\ , \label{eq:defh1}\\
h_{2}(\xieta,Q^{2}) & = & \int_{\xieta}^{1}du\ \frac{F_{2}^{(0)}(u,Q^{2})}{u^{2}}\ , \label{eq:defh2}
\end{eqnarray}
\begin{eqnarray}
h_{3}(\xieta,Q^{2}) & = & \int_{\xieta}^{1}du\ \frac{F_{3}^{(0)}(u,Q^{2})}{u}\ , \label{eq:defh3}\\
g_{2}(\xieta,Q^{2}) & = & \int_{\xieta}^{1}du\ h_{2}(u,Q^{2})
  = \int_{\xieta}^{1}du\ \int_{u}^{1}dv\ \frac{F_{2}^{(0)}(v,Q^{2})}{v^{2}}\label{eq:defg2}\\
 & = & \int_{\xieta}^{1}dv\ \int_{\xieta}^{u_{{\rm max}}=v}du\
\frac{F_{2}^{(0)}(v,Q^{2})}{v^{2}}=\int_{\xieta}^{1}dv\ (v-\xieta)\frac{F_{2}^{(0)}(v,Q^{2})}{v^{2}}\ . 
\nonumber
\end{eqnarray}
Evaluating the derivatives in Eqs.\ (\ref{eq:partf1})--(\ref{eq:partf3})
yields our final set of ``master equations'' for the target
mass corrected DIS structure functions, which we discuss next.

\subsection{\noindent Master Equations}

Combining the results in the previous section, the full, target mass corrected
structure functions can be related to the massless limit functions by the following
``master formula'', using the notation of Kretzer \& Reno (see Eq.~(3.17) of
Ref.~\cite{Kretzer:2003iu}, with their $\rho \to r$):
\begin{eqnarray}
F_{j}^{{\rm TMC}}(x,Q^{2})
&=&
\sum_{i=1}^{5}A_{j}^{i}F_{i}^{(0)}(\xieta,Q^{2})+B_{j}^{i}h_{i}(\xieta,Q^{2})
+C_{j}g_{2}(\xieta,Q^{2})\,,\  j=1-5
\ ,
\nonumber \\
\label{eq:master}
\end{eqnarray}
where $\xieta$ is the Nachtmann scaling variable from Eq.~(\ref{eq:xi})
\cite{Nachtmann:1973mr}.
Inserting the coefficients $A_{j}^{i}$, $B_{j}^{i}$, $C_{j}$ given in Tables I,II and III
of Ref.~\cite{Kretzer:2003iu}, one finds for the first three structure functions:
\begin{eqnarray}
\label{eq:master1}
F_{1}^{{\rm TMC}}(x,Q^{2}) & = & \frac{x}{\xieta r}F_{1}^{(0)}(\xieta,Q^2)
+\frac{M^{2}x^{2}}{Q^{2}r^{2}}h_{2}(\xieta,Q^2)+\frac{2M^{4}x^{3}}{Q^{4}r^{3}}g_{2}(\xieta,Q^2)\ ,
\label{eq:f1}\\
F_{2}^{{\rm TMC}}(x,Q^{2}) & = & \frac{x^{2}}{\xieta^{2}r^{3}}F_{2}^{(0)}(\xieta,Q^2)
+\frac{6M^{2}x^{3}}{Q^{2}r^{4}}h_{2}(\xieta,Q^2)+\frac{12M^{4}x^{4}}{Q^{4}r^{5}}g_{2}(\xieta,Q^2)\ , \nonumber \\
\label{eq:f2}\\
F_{3}^{{\rm TMC}}(x,Q^{2}) & = & \frac{x}{\xieta r^{2}}F_{3}^{(0)}(\xieta,Q^2)
+\frac{2M^{2}x^{2}}{Q^{2}r^{3}}h_{3}(\xieta,Q^2)+0\ ,
\label{eq:f3}
\end{eqnarray}
with the functions $h_i(\xi,Q^2)$ and $g_2(\xi,Q^2)$ 
given in Eqs.\ (\ref{eq:defh1})--(\ref{eq:defg2}).
The $F_{j}^{(0)}$ are the structure functions $F_{j}^{\rm TMC}$
in the limit $M\to0$:
\begin{equation}
F_{j}^{(0)}(\xieta,Q^{2}) \equiv
\left. \left( \lim_{M\to0}F_{j}^{\rm TMC}(x,Q^{2}) \right) \right|_{x=\xieta}\,.
\label{eq:M0limit}
\end{equation}
Note that since $\xieta$ depends on $x$ and $M$,
$F_{j}^{(0)}(\xieta,Q^2) \ne \lim_{M\to0}F_{j}^{TMC}(\xieta,Q^2)$,
which has been the source of some confusion in the literature.
Parton model representations of $F_j^{(0)}$ will be shown in Sec.\ \ref{sec:charm}.

%%%%%%%%%%%%%%%%%%%%%%%%%%%%%%%%%%%%%%%%%%%%%%%%%%%%%%%%%%%%%%%%%%%%%%%

We emphasize that the functions $F_{i}^{{\rm TMC}}=F_{i}^{{\rm TMC}}(x,Q^{2})$,
and not $F_{i}^{{\rm TMC}}=F_{i}^{{\rm TMC}}(\xieta,Q^{2})$, so that $(x,Q^{2})$
is the correct point in phase space.  While on the surface it may appear strange
to have the left-hand-side of Eq.~(\ref{eq:master}) be a function of $x$ and the
right-hand-side a function of $\xieta$, this arises quite naturally in the calculation.
Specifically, evaluating the final state momentum conservation constraint, we can
write (schematically) $\delta^{4}(q+P-P_{X})\sim\delta(x-\xieta)$, and thus
$F_{i}^{\rm TMC}(x,Q^{2}) \sim F_{i}^{(0)}(x,Q^2)\, \delta(x-\xieta)\sim F_{i}^{(0)}(\xieta,Q^2)$.
Note that it would be incorrect to write
$F_{i}^{\rm TMC}(\xieta,Q^{2}) \sim F_{i}^{(0)}(\xieta,Q^2)$.
All structure functions and PDFs depend on $Q^2$; we sometimes suppress this dependence 
for ease of notation. 

Another feature of Eq.~(\ref{eq:master}) is that $h_2$ and $g_2$ appear in the
formulas for both $F_1^{\TMC}$ and $F_2^{\TMC}$.  This follows directly from
the form of Eq.~(\ref{eq:tmunu}).  For example, both the terms proportional to
$C_1^{2k}$ and to $C_2^{2k}$ contribute to $T_1$ (multiplying $-g_{\mu\nu}$).
The terms proportional to $C_2^{2k}$ give rise to the second and third terms in
Eq.~(\ref{eq:master1}).

The ``master equation'' (\ref{eq:master}) holds to any order in the strong coupling
constant $\alpha_{s}$, which implies that the coefficients $A_{j}^{i}$, $B_{j}^{i}$ and
$C_{j}$ and the variable $\xieta$ are independent of the order (LO, NLO, NNLO, \ldots{})
to which the structure functions $F_{i}^{(0)}$ are considered. 
In addition, Eq.~(\ref{eq:master}) does not assume or imply any Callan--Gross relation.
Specifically, one can compute the longitudinal structure function according to:
\begin{eqnarray}
F_{L}^{{\rm TMC}}(x,Q^{2}) 
& = & r^{2}F_{2}^{{\rm TMC}}(x,Q^{2})
  - 2xF_{1}^{{\rm TMC}}(x,Q^{2})
\nonumber\\
& = & \frac{x^{2}}{\xieta^{2}r} \left[F_{2}^{(0)}(\xieta)-2\xieta F_{1}^{(0)}(\xieta)\right]
  +\frac{4M^{2}x^{3}}{Q^{2}r^{2}}h_{2}(\xieta)
  +\frac{8M^{4}x^{4}}{Q^{4}r^{3}}g_{2}(\xieta)
\nonumber\\
& = & \frac{x^{2}}{\xieta^{2}r}F_{L}^{(0)}(\xieta)
  +\frac{4M^{2}x^{3}}{Q^{2}r^{2}}h_{2}(\xieta)
  +\frac{8M^{4}x^{4}}{Q^{4}r^{3}}g_{2}(\xieta) \ .
\label{eq:fl}
\end{eqnarray}
This general result gives a non-zero $F_L^{\TMC}$, and thus violates the
Callan--Gross relation. 
The leading term $(\propto F_{L}^{(0)})$ is non-zero for finite quark masses, and
the sub-leading terms $(\propto h_{2},g_{2})$ contribute for finite hadron mass $M$.

%=================================================================

Final state quark mass effects are taken into account by  the
parton model structure functions $F_{i}^{(0)}$, and the general
form of the master equation is unaltered. 
In other words, the $\{A_j^i, B_j^i, C_j^i\}$ 
coefficients and the Nachtmann variable $\xi$ in Eq.~(\ref{eq:master})
will depend only on the hadronic mass and will {\it not} receive 
corrections due to final state quark masses.
We illustrate this feature for the case of neutrino production of charm quarks
in Sec.~\ref{sec:charm}, and find agreement with results in the literature
({\it cf.} \ref{app:compare}).
In particular, the proper slow-rescaling variables (which depend on the quark masses)
automatically appear as arguments of the parton distribution functions
in a natural manner ({\it cf.} \ref{app:kin}).
For the general case of non-vanishing initial and final state quark
masses we {\em assume}  the same pattern holds true---the quark masses appear
within the $F_{i}^{(0)}$ and the form of the master equation remains
unchanged; this is a consequence of factorization. 
This ensures the correct massive parton model expressions 
are, by construction, recovered in the $M \to 0$ limit; a necessary
condition for any formalism which includes target mass effects.
The modular structure of the master equation renders computations
of target mass corrections simple and transparent once the parton model
expressions for the $F_i^{(0)}$ 
(with or without quark masses, in leading or higher order)
are given.

\subsection{Nachtmann Moments}

An alternative, but closely related, formulation of the TMCs is in terms of
{\em Nachtmann moments}.  Nachtmann \cite{Nachtmann:1973mr} showed
that one could arrange the OPE so as to ensure that at a given order
in $1/Q^2$ only operators of a given twist would appear.
The Nachtmann moments $\mu_i^n$ of structure functions $F_i^{\rm TMC}$
($i = 1, 2, 3$) are constructed from operators of definite spin.
This means that from the infinite set of operators of twist-two and different
spin contained in the trace terms of the OPE, only the operators of spin $n$
contribute for the $n-2$ Nachtmann moment of the structure function.
This is contrasted with the Cornwall--Norton moments in which different spin
operators contribute to the twist-two moment.  The Nachtmann moments
are defined to factor out the target mass dependence of the structure
functions in a way such that its (Cornwall--Norton) moments would equal
the moments of the corresponding parton distributions.

For the $F_2^{\rm TMC}$ structure function for example, the Nachtmann
moment is constructed so that it depends only on the reduced matrix
element $A_n$ (and Wilson coefficient $C_i^n$) on the right-hand-side
of Eq.~(\ref{eq:f2moment}), in contrast to the infinite series of
$(M^2/Q^2)^j$ terms in Eq.~(\ref{eq:f2moment}) for the Cornwall--Norton
moment.  In effect, the subleading $(M^2/Q^2)^j$ terms with $j \geq 1$
are absorbed in the redefined moment on the left-hand-side of
Eq.~(\ref{eq:f2moment}), so that only the first term with $j=0$
contributes to the Nachtmann moment.

Specifically, the Nachtmann moment of the $F_2$ structure function is
given by:
\begin{eqnarray}
\mu_2^n(Q^2)
&=& \int_0^1 dx {\xieta^{n+1} \over x^3}
    \left[ { 3 + 3(n+1)r + n(n+2)r^2 \over (n+2)(n+3) }
    \right]
    F_2^{\rm TMC}(x,Q^2)\ ,
\label{eq:mu_2}
\end{eqnarray}
where again we use $r = \sqrt{1 + 4 x^2 M^2/Q^2}$.
One can also express the Nachtmann moment as an integral over the
Nachtmann variable $\xieta$,
\begin{eqnarray}
\nonumber
\mu_2^n(Q^2)
&=& \int_0^{\xieta_0} d\xieta\ \xieta^{n-2}\
   \frac{(1 + \xieta^2 M^2/Q^2)^3}{1 - \xieta^2 M^2/Q^2}\	
   \left[1 - \frac{3(r - 1)}{r^2(n+2)} - \frac{3(r-1)^2}{r^2(n+3)}
   \right]\ \\
  &\times\ & F_2^{\rm TMC}(x,Q^2)\mid_{x=\xieta/(1-M^2\xieta^2/Q^2)},
\end{eqnarray}
with $F_2^{\rm TMC}(x,Q^2)$ given by Eq.~(\ref{eq:f2}), and
$\xieta_0 = \xieta(x=1) = 2 / (1 + \sqrt{1 + 4 M^2/Q^2})$.

Similarly, for the longitudinal Nachtmann moments (or equivalently for the
$F_1^{\rm TMC}$ structure function, defined in terms of $F_2^{\rm TMC}$
and $F_L^{\rm TMC}$), one has \cite{Nachtmann:1973mr}:
\begin{eqnarray}
\mu_L^n(Q^2)
&=& \int_0^1 dx\ { \xieta^{n+1} \over x^3 }
\Biggl\{
  F_L^{\rm TMC}(x,Q^2) \nonumber \\ 
&+&  {4 M^2 x^2 \over Q^2} { (n+1) \xieta/x - 2 (n+2) \over (n+2)(n+3) }
  F_2^{\rm TMC}(x,Q^2)
\Biggr\} \ .
\label{eq:NachtLMom}
\end{eqnarray}
Note that while in the $Q^2 \to \infty$ limit $\mu_L^n(Q^2)$ approaches
the Cornwall--Norton moment of $F_L^{\rm TMC}$, at finite $Q^2$ both
$F_L^{\rm TMC}$ and $F_2^{\rm TMC}$ contribute.

The Nachtmann moment $\mu_2^n(Q^2)$ and Cornwall--Norton moment
$M_2^n(Q^2)$ can be related by expanding the moments in powers of
$1/Q^2$.  Expanding $\mu_2^n$ to ${\cal O}(1/Q^6)$, one has:
\begin{eqnarray}
\mu_2^n(Q^2)
&=& M_2^n(Q^2)\ -\ { n(n-1) \over n+2 } {M^2 \over Q^2} M_2^{n+2}(Q^2)\
\nonumber \\
& +& \ {n(n^2 - 1) \over 2 (n+3)} {M^4 \over Q^4} M_2^{n+4}(Q^2)
 -  {n (n^2 - 1) \over 6} {M^6 \over Q^6} M_2^{n+6}\
 +\ \cdots\ ,
\label{eq:cnmoments}
\end{eqnarray}
which illustrates the mixing between the lower and higher
Cornwall--Norton moments.

%---------------------------------------------
\begin{figure}[!t]
\begin{center}
\includegraphics[scale=0.60]{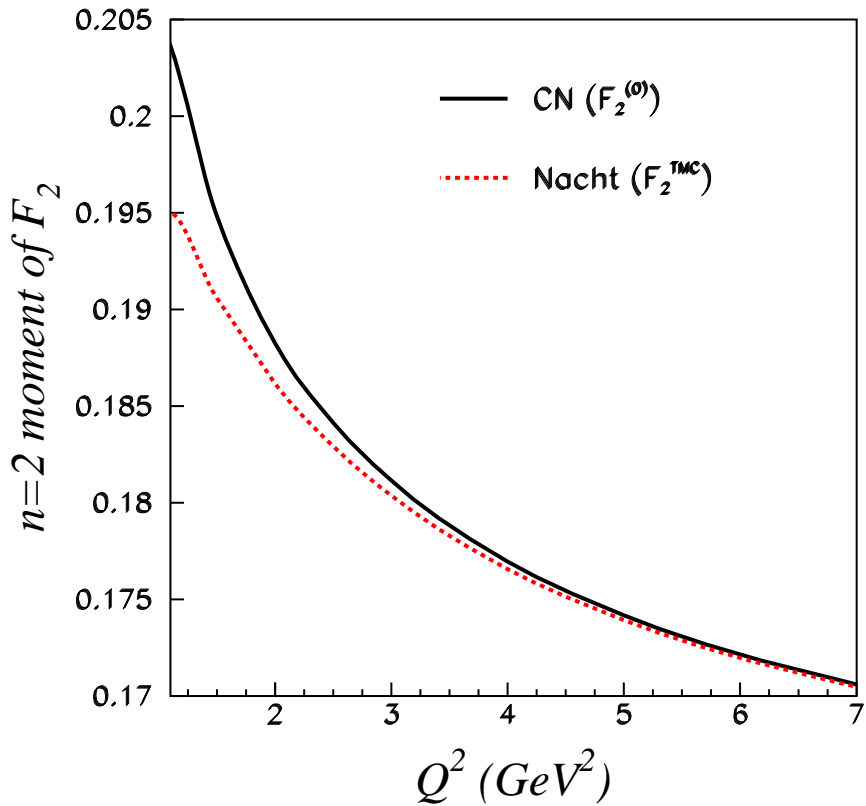}
\caption{Comparison of the $n=2$ Cornwall--Norton moment of the proton
	$F_2^{(0)}$ structure function (solid), with the Nachtmann moment of
	the target mass corrected $F_2^{\TMC}$ (dotted), calculated from
	$F_2^{(0)}$ using Eq.~(\ref{eq:master}).}
\label{fig:cnNacht}
\end{center}
\end{figure}
%---------------------------------------------

In Fig.~\ref{fig:cnNacht} we compare the $n=2$ Nachtmann moment
$\mu_2^{n=2}$ of the target mass corrected proton $F_2^{\TMC}$
structure function (solid curve) with the Cornwall--Norton moment
$M_2^{n=2}$ (dotted curve) of the same structure function in the
massless target limit, $F_2^{(0)}$ (note the expanded vertical scale!).
While the two moments agree well at large $Q^2$, a clear deviation
from equality is seen for $Q^2 \lsim 2~\GeV^2$.  Part of this
discrepancy may be attributed to the behavior of the target mass
corrected structure function in the $x \rightarrow 1$ limit, as we
discuss in detail in Sec.~\ref{sec:xto1}.

%%%%%%%%%%%%%%%%%%%%%%%%%%%%%%%%%%%%%%%%%%%%%%%%%%%%%%%%%%%%%%%%%%%
%%%%%%%%%%%%%%%%%%%%%%%%%%%%%%%%%%%%%%%%%%%%%%%%%%%%%%%%%%%%%%%%%%%
%%%%%%%%%%%%%%%%%%%%%%%%%%%%%%%%%%%%%%%%%%%%%%%%%%%%%%%%%%%%%%%%%%%
\section{Relation of the OPE to the Parton Model \label{sec:OPEpm}}

In the previous section we used the OPE to relate the 
 inclusive lepton--nuclear scattering 
 cross section
to the structure functions $F_{i}^{(0)}$ of Eq.~(\ref{eq:master}).
While these relations are quite general, they are of no utility unless
we can relate them to calculable quantities. 
Up to this point we have not invoked the parton model and its associated
assumptions; the only ingredient has been the OPE, which only makes
use of fundamental symmetries.
In this section we briefly discuss the relation of the leading twist OPE
treatment of inclusive lepton--nuclear scattering  with the parton model.
We also comment on the importance of TMCs in testing the validity
of leading twist descriptions of data, or alternatively, extracting higher
twist contributions.

The leading, $j=0$, term in Eq.~(\ref{eq:pi}) reduces the target mass
corrected structure function $F_i^{\TMC}$ to the massless limit function
$F_i^{(0)}$, which can be expressed in terms of parton distributions
in the parton model.  The $j>0$ terms in Eq.~(\ref{eq:pi}) resum the
target mass corrections.  The explicit form of the structure functions
$F_i^{(0)}$ depends on the interaction.  For $F_1^{(0)}$, for example,
the $j=0$ term in Eq.~(\ref{eq:cnF13}) can be schematically written
(neglecting quark electroweak charges) as:
\begin{eqnarray}
C_1^{n}\cdot A_{n} &=& \int_0^1 dx x^{n-1} F_1^{(0)}(x,Q^2)
\label{III.lt1}	\\ 
&=& \int_0^1 dz z^{n-1} \sigma(z,Q^2,\mu^2)\cdot 
	\int_0^1 dy y^{n-1} f(y,\mu^2)
\label{III.lt2}	\\
&=& \int_0^1 dx x^{n-1}
	\int_x^1 \frac{dy}{y} \sigma\Bigl(\frac{x}{y},Q^2,\mu^2\Bigr)\
	f(y,\mu^2)\ ,
\label{III.lt3}
\end{eqnarray}
where $f$ is the parton distribution function, defined at a scale $\mu$,
and $\sigma$ is the boson--parton scattering cross section.
The first factor in the product $C_1^n \cdot A_n$ above is the 
process-dependent Wilson coefficient, while the second is the
moment of the parton distribution $f$. At leading order in QCD
perturbation theory, the parton cross section
$\sigma(z,Q^2,\mu^2)\sim \delta(1-z)$, multiplied by numerical
factors that account for the process (such as $e_f^2/2$ for 
electromagnetic scattering with a parton of electric charge $e_f$).
QCD corrections introduce a $Q^2$ dependence to $\sigma$ and
therefore to $C_i^{n}$. Both $C_i^{n}$ and $A^{n}$ depend on
$\mu^2$ in such a way that the product is $\mu^2$ independent.

As remarked above, Eq.~(\ref{eq:master}) is valid at leading twist,
and to all orders in $\alpha_{s}$.  The inclusion of higher orders
in $F_i^{(0)}$ is standard in the leading twist approximation.
This involves evaluating the perturbative QCD corrections to
$Vq \rightarrow q$ and associated gluon processes. The NLO
corrections have been known for some time
\cite{Altarelli:1979ub,Gottschalk:1980rv,Furmanski:1981cw,Gluck:1987uk,Gluck:1996ve}, 
and the NNLO corrections have also been evaluated
\cite{vanNeerven:1991nn,Zijlstra:1992qd,Zijlstra:1992kj,Vogt:2004mw,Moch:2004xu,Vermaseren:2005qc}.

In addition to the perturbative corrections, moments of structure
functions in the OPE receive $1/Q^2$ power corrections, which are
suppressed at large $Q^2$, but may be significant at lower $Q^2$.
The size of these corrections is determined by matrix elements of
local operators which have twist greater than 2, denoted
``dynamical higher twist''.
Since one of the motivations for studying TMCs is to more reliably
extract leading twist parton distributions from structure function data,
it is necessary to ensure that TMCs are properly taken into account
so as to reveal genuine effects associated with dynamical higher twists.
Of course, the TMCs themselves enter as $1/Q^2$ corrections, despite
formally being of leading twist, so in practice it is crucial to remove from
the data TMC effects which could otherwise resemble higher twists.

The general expansion for the moments of structure functions in QCD is
then one which involves an expansion in powers of $1/Q^2$, each term of
which has an associated perturbative $\alpha_s$ expansion, in addition
to the $(M^2/Q^2)^j$ TMC expansion as in Eq.~(\ref{eq:f2moment}).
For the above example of the $F_1$ structure function, the general OPE
expansion in the $M \to 0$ limit can be written \cite{Miramontes:1988fz}:
\begin{eqnarray}
\label{eq:highertwist1}
\int_0^1 dx\ x^{n-1} F_1^{(0)}(x,Q^2)\
&=&
\ C_1^n(Q^2)\ A_n\
+\ \sum_{i=1}^\infty C_1^{n,i}(Q^2)\ \Biggl(\frac{n M_0^2}{Q^2}\Biggr)^i\
  B_{n, i}\ ,
\nonumber \\
\end{eqnarray}
where the first (twist-two) term corresponds to that in Eq.~(\ref{III.lt1}),
and the higher twist coefficients $B_{n, i}$ are of the same order as
$A_n$, with $M_0$ some typical hadronic mass scale.
Several analyses of low $Q^2$ data have been made which include
both leading and higher twist contributions, although the practical
difficulty in disentangling $1/Q^2$ from even higher order corrections
means that usually terms only up to twist 4 are considered.

\begin{figure}[!t]
\begin{center}
\includegraphics[scale=0.40]{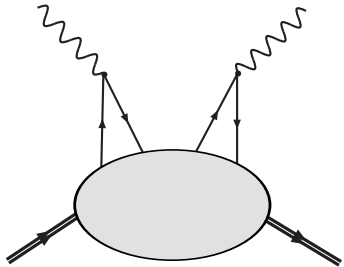}
\caption{
An example of a higher twist diagram, involving a four-quark
operator, contributing to the term $B_{n,i}$ in
Eq.~(\ref{eq:highertwist1}).
}
\label{fig:catears} 
\end{center}
\end{figure}

Although in the context of parton distribution analyses higher twists
are seen as unwelcome complications, the study of higher twists is also
important in its own right.
Because the higher twist operators necessarily involve several quark
or quark and gluon fields, their matrix elements can provide insight into
nonperturbative, multi-parton correlations in the nucleon, and possibly
even on the long distance partonic interactions associated with
confinement.
For example, twist-4 corrections to the $F_2$ structure function
involve the so-called ``cat's ears'' diagrams schematically shown in 
Fig.~\ref{fig:catears}, which represent flavor non-diagonal transitions
between the incoming and outgoing photons in the Compton scattering
process \cite{Ji:1994br}.
Differences in their magnitude in the proton and neutron can reveal
the mechanism responsible for the phenomenon of Bloom--Gilman duality
in structure functions \cite{Melnitchouk:2005zr}.
For spin dependent structure functions, certain twist-3 and twist-4
matrix elements can be related to color polarizabilities of the nucleon,
which describe how the color electric and magnetic gluon fields respond
to the spin of the nucleon
\cite{Shuryak:1981pi,Stein:1995si,Osipenko:2004xg,Meziani:2004ne}.
Before we can begin to unravel these effects, it is essential that
the kinematical TMCs are properly removed from the data.

%--------------------------------------------------------------------
%--------------------------------------------------------------------

Having outlined the formalism, in the next section we display explicit
expressions for $F_{i}^{(0)}$ for leading order QCD at leading twist
for neutrino--proton scattering.  We focus on this process because of
the subtleties associated with charm production by massless quarks.
Next-to-leading order expressions for the structure functions $F_{i}^{(0)}$
in a fixed flavor number scheme with three active flavors (3-FFNS) can
be found in Ref.~\cite{Kretzer:2002fr}.

%%%%%%%%%%%%%%%%%%%%%%%%%%%%%%%%%%%%%%%%%%%%%%%%%%%%%%%%%%%%%%%%%%%
%%%%%%%%%%%%%%%%%%%%%%%%%%%%%%%%%%%%%%%%%%%%%%%%%%%%%%%%%%%%%%%%%%%
%%%%%%%%%%%%%%%%%%%%%%%%%%%%%%%%%%%%%%%%%%%%%%%%%%%%%%%%%%%%%%%%%%%
\section{Corrections for Finite Hadron and Quark Masses\label{sec:charm}}

\begin{figure}[!t]
\begin{center}
\includegraphics[scale=0.65]{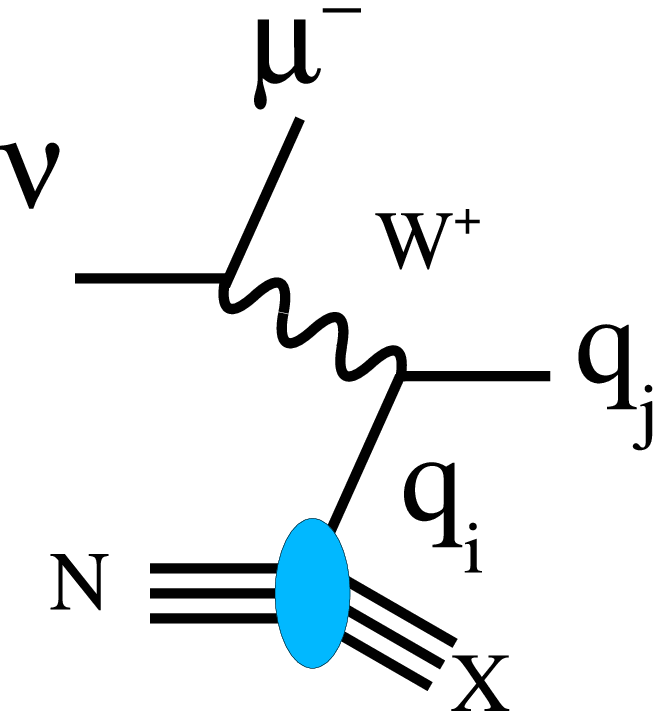}
\hfil
\includegraphics[scale=0.65]{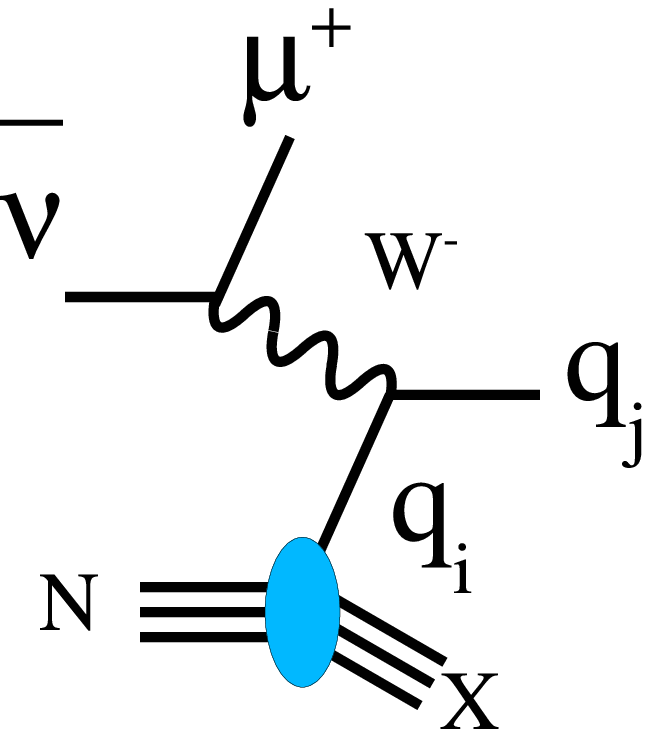}
\caption{
The basic Feynman diagrams for the DIS process for 
neutrinos, $\nu N \to \mu^- X$ (left diagram) and 
anti-neutrinos $\bar\nu N \to \mu^+ X$ (right diagram).
}
\label{graph} 
\end{center}
\end{figure}

To illustrate how the OPE--parton model formalism is implemented in
the case of a quark mass scale $m_j$ in the final state (in addition to the target mass
scale $M$), we consider the case of neutrino inclusive lepton--nuclear scattering 
 for 3 light $\{u,d,s\}$
flavors and one heavy $\{c\}$ flavor. 
The reaction takes place via the process  $\nu_\mu N \to  \mu^- X$
(or $\bar\nu_\mu N \to  \mu^+ X$), {\it cf.} Fig.~\ref{graph}.
In this example we will focus on the leading order only, so we can
separately observe how both $M$ and $m_j=m_c$ enter; once the pattern
is developed here, the generalization to multiple quark masses ($m_i, m_j$) 
will be outlined in \ref{app:kin}.

%--------------------------------------------------------------------
%--------------------------------------------------------------------

\subsection{Multiple Mass Scales in Neutrino DIS}

To illustrate how the quark and hadron masses enter into the structure
functions, we separate the $M \to 0$ limit structure functions into
contributions with a light ($l$) and a heavy quark ($c$) in the final state.
For neutrino scattering, we thus have:
\begin{equation}
F_{i}^{\nu,(0)}(\xi ,Q^{2})
=
F_{i,l}^{\nu,(0)}(\xi,Q^{2})
+
F_{i,c}^{\nu,(0)}(\xi,Q^{2})\ .
\label{eq:Fnu}
\end{equation}
For neutrino DIS with a light quark in the final state ({\em i.e.}, no charm),
the structure functions for a 3-flavor number scheme are given by:
\begin{eqnarray}
F_{1,l}^{{\nu}(0)}(\xi,Q^{2})
& = & d(\xi) |V_{ud}|^{2}+s(\xi)|V_{us}|^{2}
	+\bar{u}(\xi)(|V_{ud}|^{2}+|V_{us}|^{2})\ , 
\label{eq:f1l}
\\
F_{2,l}^{\nu,(0)}(\xi,Q^{2})
& = & 2\xi\ \left[d(\xi)|V_{ud}|^{2}+s(\xi)|V_{us}|^{2}
	+\bar{u}(\xi)(|V_{ud}|^{2}+|V_{us}|^{2})\right]\ , 
\label{eq:f2l}
\\
F_{3,l}^{\nu,(0)}(\xi,Q^{2})
& = & 2\left[d(\xi)|V_{ud}|^{2}+s(\xi)|V_{us}|^{2}
	- \bar{u}(\xi)(|V_{ud}|^{2}+|V_{us}|^{2})\right]\ ,
\label{eq:f3l}
\end{eqnarray}
where $V_{ij}$ are the CKM matrix elements, and the $Q^{2}$
dependence in the PDFs has been suppressed.
For heavy quarks, the structure function contributions are:
\begin{eqnarray}
\label{eq:f1c}
F_{1,c}^{{\nu}(0)}(\xi,Q^{2})
& = & d(\bar{\xi})|V_{cd}|^{2}+s(\bar{\xi})|V_{cs}|^{2}\ , \\
F_{2,c}^{\nu,(0)}(\xi,Q^{2})
& = & 2\bar{\xi}\left[d(\bar{\xi})|V_{cd}|^{2}+s(\bar{\xi})|V_{cs}|^{2}\right]
\label{eq:f2c}\ ,\\
F_{3,c}^{\nu,(0)}(\xi,Q^{2})
& = & 2\left[d(\bar{\xi})|V_{cd}|^{2}+s(\bar{\xi}) |V_{cs}|^{2}\right]\ .
\label{eq:f3c}
\end{eqnarray}
For anti-neutrino structure functions $F_{i}^{\bar\nu,(0)}(\xi ,Q^{2})$, the results
are obtained by interchanging $q \leftrightarrow \bar q$ in Eq.~(\ref{eq:Fnu}):
\begin{equation}
F_{1,2}^{\bar{\nu},(0)}=F_{1,2}^{\nu,(0)}[q\leftrightarrow\bar{q}],
\quad
F_{3}^{\bar{\nu},(0)}=-F_{3}^{\nu,(0)}[q\leftrightarrow\bar{q}],
\quad q=u,d,s\,.
\end{equation}

The variable $\bar{\xi}$ in Eqs.~(\ref{eq:f1c})--(\ref{eq:f3c}) is the so-called
slow rescaling variable, generalized to include both target and quark mass
effects, and corresponds to the light-cone fractional momentum of a massless
quark which produces a charm quark.  It is related to the Nachtmann scaling
variable $\xi$ by $\bar{\xi} \equiv \xi R_{ij}$, where $R_{ij}$ contains the
dependence on the initial and final quark masses ({\em cf.} Eq.\ (\ref{eq:Rij})), 
and in the limit of a single
heavy quark $j=\{c\}$, and three light flavors $i=\{u,d,s\}$ in the initial state, 
it reduces to $R_{ij}=(1+m_{c}^{2}/Q^{2})$.
In this case we recover the relation: 
$\bar{\xi} = \xi\ (1+m_{c}^{2}/Q^{2})$ \cite{Barnett:1976ak,Aivazis:1993kh,Aivazis:1993pi}.

Note that Eqs.~(\ref{eq:f1l}) and (\ref{eq:f2l}) lead to a generalized Callan--Gross
relation at leading-order (with $x$ replaced by $\xi$):
$2 \xi F_{1,l}^{\nu,(0)}(\xi) = F_{2,l}^{\nu,(0)}(\xi)$.
It is also important to note in Eqs.~(\ref{eq:f1c}), (\ref{eq:f2c}), and (\ref{eq:f3c}) 
that operationally $F_i^{{\nu}(0)}$ is a function of the variable $\xi$, while
the PDFs are functions of $\bar\xi = \xi R_{ij}$.  As discussed in \ref{app:charm},
this is simply a consequence of evaluating the final state partonic momentum
conserving delta-function $\delta(\xi - \bar\xi /R_{ij})$.

\subsection{Relation to TMC Structure Functions} 

From the master equation (\ref{eq:master}), we can relate the above
partonic distributions to the TMC structure functions. The  charm mass
effects can be taken into account in the structure functions $F_{i}^{{\nu}(0)}$
in Eqs.~(\ref{eq:f1c})--(\ref{eq:f3c}) by using the generalized Nachtmann
variable $\bar\xi$, thereby allowing the form of Eq.~(\ref{eq:master})
to be kept unmodified.  For example, in the limit of a diagonal CKM matrix,
$V_{ij}=\delta_{ij}$, and neglecting the non-leading terms, inserting
$F_{2,c}^{\nu,(0)}(\xi,Q^{2})\simeq2\bar{\xi}s(\bar{\xi})$ into the leading
term of Eq.~(\ref{eq:f2}) gives: 
\begin{eqnarray}
F_{2,c}^{{\rm TMC}}(x,Q^{2})
&\simeq& \frac{x^{2}}{\xi^{2}r^{3}}\ F_{2,c}^{\nu,(0)}(\xi,Q^{2})
\simeq   \frac{x^{2}}{\xi^{2}r^{3}}\ 2\bar{\xi}s(\bar{\xi}) 
\nonumber\\
&=& \frac{2x^{2}}{r^{3}}\ (1+m_{c}^{2}/Q^{2})^{2}\ \frac{s(\bar{\xi},Q^{2})}{\bar{\xi}}\ .
\end{eqnarray}

The complete set of expressions in this ``diagonal CKM'' limit for the charm
production component of the target mass corrected structure functions is
given by:
\begin{eqnarray}
F_{1,c}^{{\rm TMC}}(x,Q^{2}) 
& \simeq & 
\frac{x}{\bar{\xi}r}(1+m_{c}^{2}/Q^{2})s(\bar{\xi})
+\frac{2M^{2}x^{2}}{Q^{2}r^{2}}(1+m_{c}^{2}/Q^{2})\int_{\bar{\xi}}^{1}\der u\ \frac{s(u)}{u}\ ,
\nonumber\\
&  & +\frac{4M^{4}x^{3}}{Q^{4}r^{3}}\int_{\bar{\xi}}^{1}\der u\ \int_{u}^{1}\der v\ \frac{s(v)}{v}
\label{eq:f1lo}\ , \\
F_{2,c}^{{\rm TMC}}(x,Q^{2}) 
& \simeq & 
\frac{2x^{2}}{\bar{\xi}r^{3}}(1+m_{c}^{2}/Q^{2})^{2}s(\bar{\xi})
+\frac{12M^{2}x^{3}}{Q^{2}r^{4}}(1+m_{c}^{2}/Q^{2})\int_{\bar{\xi}}^{1}\der u\ \frac{s(u)}{u}
\nonumber\\
 &  & +\frac{24M^{4}x^{4}}{Q^{4}r^{5}}\int_{\bar{\xi}}^{1}\der u\ \int_{u}^{1}\der v\ \frac{s(v)}{v}
\label{eq:f2lo}\ , \\
F_{3,c}^{{\rm TMC}}(x,Q^{2}) 
& \simeq & 
\frac{x}{\bar{\xi}r^{2}}(1+m_{c}^{2}/Q^{2})2s(\bar{\xi})
+\frac{2M^{2}x^{2}}{Q^{2}r^{3}}\int_{\bar{\xi}}^{1}\der u\ \frac{2s(u)}{u}+0\ .
\label{eq:f3lo}
\end{eqnarray}
A discussion of the dependence of $h_i$ and $g_2$ on the charm
quark mass is given in \ref{app:charm}.
\footnote{See also Ref.~\cite{Kretzer:2003iu}, Eqs.~(3.1)--(3.3);
with $g_2$ and $h_3$ defined as in Eqs.~(\ref{eq:g2lo}) and (\ref{eq:h3lo}), the
charm mass dependent structure functions of Eqs.~(\ref{eq:f1lo})--(\ref{eq:f3lo})
are recovered.}

%%%%%%%%%%%%%%%%%%%%%%%%%%%%%%%%%%%%%%%%%%%%%%%%%%%%%%%%%%%%%%%%%%%%%%%
%%%%%%%%%%%%%%%%%%%%%%%%%%%%%%%%%%%%%%%%%%%%%%%%%%%%%%%%%%%%%%%%%%%%%%%

Having illustrated the implementation for non-zero hadron and quark
masses, the extension to the general mass case is straightforward
from the relations in \ref{app:charm}.  As with the above charm mass
case, the form of the equations will remain the same; the target mass
dependence enters in the replacement $x \to \xi = x R_M$, where
$R_M = 2/(1+r)$ (see \ref{app:kin}), and the quark masses enter via
the replacement $\xi \to \bar\xi = \xi R_{ij}$.

%%%%%%%%%%%%%%%%%%%%%%%%%%%%%%%%%%%%%%%%%%%%%%%%%%%%%%%%%%%%%%%%%%%%%%%%%
\section{Threshold Effects and the $x\to1$ Limit
\label{sec:xto1}}

The standard TMC formulation described in the previous sections
involves the so-called ``threshold problem'', which was recognized
soon after the original derivation of TMCs, and has been discussed
by a number of authors in the literature 
\cite{Gross:1976xt,DeRujula:1976ih,Bitar:1978cj,Johnson:1979ty,Steffens:2006ds}.
It is associated with the behavior of parton distributions in the
threshold region, between pion-production ($W = M + m_\pi$) and the
elastic point ($W = M$), and becomes increasingly important as $x \to 1$.

The problem can be summarized as follows. For simplicity, we neglect
perturbative QCD (pQCD) corrections, with the Wilson coefficient
functions $C_i^n$ set to unity. Following Ref. \cite{Georgi:1976ve},
we define a leading twist parton distribution in this discussion by
$F(\xi,Q^2) \equiv F_2^{(0)}(\xi,Q^2)/\xi^2$ ({\em cf.} Eq.~(\ref{eq:cnF2})).
The $n$-th moment of the parton distribution,
\begin{eqnarray}
A_n &=& \int_0^1 d\xi\ \xi^n\ F(\xi,Q^2)\ ,
\end{eqnarray}
should be $Q^2$ independent.
Neglecting higher twist (HT) contributions, one should have
\cite{DeRujula:1976ih,Gross:1976xt}:
\begin{equation}
\int_0^1 d\xi\ \xi^{n-2}\ F_2^{(0)}(\xi,Q_1^2)
= \int_0^1 d\xi\ \xi^{n-2}\ F_2^{(0)}(\xi,Q_2^2)\quad
{\rm [no\ pQCD,\ no\ HT]}
\label{eq:anequality}
\end{equation}
for any two momentum scales $Q_1^2$ and $Q_2^2$, where
$F_2^{(0)}(\xi,Q^2)$ is the structure function in the massless
target limit, $M \to 0$ (see Eq.~(\ref{eq:M0limit})).
Since $F_2^{(0)}(\xi,Q^2)$ must vanish in the kinematically
forbidden region $\xi > \xi_0$, where
\begin{equation}
    \xi_0 (Q^2)\equiv \xi(x=1,Q^2) = 2/(1+\sqrt{1+4M^2/Q^2})\ ,
\end{equation}
the equality in Eq.~(\ref{eq:anequality})
implies that $F_2^{(0)}(\xi,Q^2)$ must be zero for both $\xi > \xi_0(Q_1^2)$
and $\xi > \xi_0(Q_2^2)$.
If $Q_1^2 < Q_2^2$, in which case $\xi_0(Q_1^2) < \xi_0(Q_2^2)$,
this implies that $F_2^{(0)}(\xi,Q_2^2)$ must vanish in the range
$\xi_0(Q_1^2) < \xi < \xi_0(Q_2^2)$.
However, there is no physical reason in the parton
model for it to do so here, and
this leads to an unphysical constraint.

According to the prescription of De R\'ujula, Georgi and Politzer (DGP)
\cite{DeRujula:1976ih}, the solution to this ``paradox'' lies in the
higher order terms in the twist expansion for the moments of
$F_2^{(0)}(\xi,Q^2)$.
In general, the $n$-th moment of the structure function
$F_2^{(0)}(\xi,Q^2)$ can be written as \cite{DeRujula:1976ih}
({\em cf.} Eq.~(\ref{eq:highertwist1}))
\begin{eqnarray}
\int_0^1 d\xi\ \xi^{n-2}\ F_2^{(0)}(\xi,Q^2)
&=& A_n\
 +\ \sum_{i=1}^\infty \left( { n M_0^2 \over Q^2 } \right)^i\
    B_{n,i}\ ,
\label{eq:OPE}
\end{eqnarray}
where the $i$-th term in the sum is the contribution of operators
of twist $2i+2$.

DGP argue \cite{DeRujula:1976ih} that the higher twist $B_{n,i}$ terms
indeed tame the paradox.  They note that if the distribution at large $\xi$
behaves as $(1-\xi)^a$, with $a \approx 3-4$, then the $n$-th moment
of $F_2^{(0)}$ will be sensitive to the distribution at values of $\xi$ close
to $\xi_n \equiv n/(n+a) \approx 1 - a/n$ for large $n$.  For the moments
to be sensitive to the threshold region, one must take $n$ large enough
so that $\xi_n \approx \xi_0(Q^2) \approx 1 - M^2/Q^2$ at large $Q^2$.
In other words, the analysis is affected strongly by the region in moment
space where $n \approx a Q^2/M^2$.  For such large values of $n$,
however, the terms proportional to $B_{n,i}$ cannot be disregarded,
and DGP conclude that there is therefore no paradox.

More specifically, there is a non-uniformity in the limits $n \to \infty$
(or $\xi \to 1$) and $Q^2 \to \infty$. The appearance of powers of
$n M_0^2/Q^2$ in Eq.~(\ref{eq:OPE}) signals the breakdown of the
whole twist expansion in this region.  The master equation (\ref{eq:master})
is therefore not valid for large $\xi$, where higher twists are important.

The problem with the nonuniform limits was also recognized by
Tung and collaborators \cite{Bitar:1978cj,Johnson:1979ty}, in the
context of Nachtmann moments, which were discussed in Sec.\ \ref{sec:OPE}.
In particular, they pointed out that the Nachtmann moments
$\mu_i^n(Q^2)$ of structure functions have, for fixed $n$, the
asymptotic $Q^2 \to \infty$ limit (again, neglecting pQCD corrections) 
\begin{eqnarray}
\mu_i^n(Q^2)
&\to& 
      A_n\ \ \ \ \
[n\ {\rm fixed},\ Q^2 \to \infty, \ {\rm no\ pQCD}]\ .
\label{eq:mu_pQCD}
\end{eqnarray}
However, the asymptotic form does not exhibit the correct threshold
behavior at fixed $Q^2$, which should be
\cite{Bitar:1978cj,Johnson:1979ty}
\begin{eqnarray}
\mu_i^n(Q^2)
&\to& \xi_0^n(Q^2)\ \widetilde\mu_i^n(Q^2)\ \ \ \ \ \ \
[Q^2\ {\rm fixed},\ n \to \infty]\ ,
\label{eq:mu_eta0}
\end{eqnarray}
where $\widetilde\mu_i^n(Q^2)$ are the ``regularized'' moments
of a function which has support over the correct physical range,
$0<\xi<\xi_0$. In the asymptotic $n\to \infty$ limit, the regularized
moments are then weakly dependent upon $n$, while the full
moments $\mu_i^n$ contain the main $n$ dependence through
the threshold factor $\xi_0^n(Q^2)$ \cite{Bitar:1978cj,Johnson:1979ty}.

Tung {\em et al.} \cite{Bitar:1978cj,Johnson:1979ty} proposed a 
solution to this problem by postulating an {\em ansatz} for the
moments in which the threshold factor is explicitly included
in the definition,
\begin{eqnarray}
\mu_i^{n\ (\rm Tung)}(Q^2)
&\equiv& \xi_0^n(Q^2)\
	 A_n\ ,
\label{eq:mu_Tung}
\end{eqnarray}
where the new moments are consistent with both the asymptotic QCD
behavior (\ref{eq:mu_pQCD}) and the kinematic threshold requirement
(\ref{eq:mu_eta0}).
However, as they note, such a prescription is not unique, and in fact
agrees with the standard OPE expansion only in the $n \to \infty$ limit
\cite{Steffens:2006ds}.

The region of $Q^2$ where TMCs are significant also corresponds
to the region where nucleon resonances play an important role,
$W \lesssim 2$~GeV.
Although the resonance region displays significant $Q^2$ dependent
structure, a twist expansion is still useful here when one averages
over individual resonances.
The resonance region corresponds to $n \lesssim a Q^2/M^2$, and here
higher twists are small but not negligible.
The size of the higher twists in fact determines the degree to which
the physical structure function oscillates around the leading twist
function, in the sense of local Bloom--Gilman duality \cite{Bloom:1970xb}.
Target mass corrections are therefore closely related to the physics
of resonances and quark--hadron duality \cite{Melnitchouk:2005zr}.

Duality is also invoked implicitly in the DGP approach in that
the parton distribution in the region between the pion production
threshold, corresponding to $W = M + m_\pi$, and $\xi=1$ is taken
to be dual to the elastic form factor at $\xi = \xi_0$ 
\cite{Bloom:1970xb,Melnitchouk:2001eh,Steffens:2004vw}.
However, a consequence of a non-zero parton distribution $F(\xi,Q^2)$
over the entire region $0 \leq \xi \leq 1$, including the unphysical
region $\xi > \xi_0$ (at $x > 1$), is that the target mass corrected 
structure function is non-zero at $x=1$, {\em i.e.},
$F_2^{\rm TMC}(x=1,Q^2) > 0$, for any finite $Q^2$.
This seems necessary in the DGP approach if one wishes to preserve a
probabilistic interpretation for $F(\xi)$, as in the parton model.

An alternative to working with distributions in the unphysical region
$\xi > \xi_0$ was suggested by Steffens and
Melnitchouk (SM) in Ref.~\cite{Steffens:2006ds}.
The philosophy adopted there was to reformulate the DGP analysis in
such a way as to ensure the correct kinematic limit for the target
mass corrected structure function, {\em i.e.},
$F_2^{\rm TMC}(x\to 1,Q^2) \to 0$, for any finite $Q^2$.
The method specifically involved working with parton distributions
which vanish beyond the kinematic upper limit, $\xi = \xi_0$, and
which at finite $Q^2$ therefore depend on $\xi$ and $\xi_0$.

The approach of SM \cite{Steffens:2006ds} defines the matrix element 
$A_n^{\rm (SM)}$ in terms of an integral of the parton distribution over the range
$0 \leq \xi \leq \xi_0$,
\begin{eqnarray}
A_n^{{\rm (SM)}} 
&\equiv& \int_0^{\xi_0} d\xi\ \xi^n\ F(\xi,\xi_0)\ ,
\end{eqnarray}
where the function $F$ is now $\xi$ and $\xi_0$ dependent,
and vanishes smoothly as $\xi \to \xi_0$, for instance as
a power of $(\xi_0 - \xi)$.
The function $F$ is also properly normalized; for example,
if the functional form for $F$ is
$F(\xi,\xi_0) = N\ \xi^a\ (\xi_0 - \xi)^b$, then the
normalization constant is $N = \xi_0^{-a-b-1}$.

One can also define a ``scaled Nachtmann variable''
\begin{eqnarray}
\tilde\xi &=& { \xi \over \xi_0 }\ ,
\end{eqnarray}
and write $A_n^{\rm (SM)}$ in a form resembling that in DGP,
\begin{eqnarray}
A_n ^{{\rm (SM)}}
&=& \xi_0^n \int_0^1 d\tilde\xi\ \tilde\xi^n\ \widetilde
F(\tilde\xi)\ ,
\label{eq:TD}
\end{eqnarray}
where now the function $\widetilde F$ depends {\em only} on a
single variable $\tilde\xi$.
In this approach, one does not need to invoke higher twists to
cancel the leading twist TMCs in the $\xi > \xi_0$ region.
In the original DGP formulation, the extended, unphysical range
$\xi > \xi_0$ implies that the target mass corrected structure
function is overestimated at large $\xi$.
A resulting complication of this is that
even neglecting pQCD, the moments $A_n^{\rm (SM)}$ become 
$Q^2$ dependent, thereby making the connection with the OPE, and
hence a partonic interpretation, problematic in the presence of TMCs.
This $Q^2$ dependence is, however, in accordance with the results
of Tung {\em et al.}, Eq.~(\ref{eq:mu_Tung}).

An alternative approach to avoid introducing this additional
$Q^2$ dependence is to redefine the moments such that
\begin{eqnarray}
A_n ^{{\rm (SM')}}&\equiv& \int_0^{\xi_0} d\xi\ \left(\frac{\xi}{\xi_0}\right)^n
F(\xi,\xi_0)\ .
\end{eqnarray}
In terms of the variable $\tilde\xi$ this can then be rewritten as
\begin{eqnarray}
A_n^{{\rm (SM')}} 
&=& \int_0^1 d\tilde\xi\ \tilde\xi^n\ \widetilde
F(\tilde\xi)\ .
\label{eq:GP}
\end{eqnarray}
This definition leads to the standard approach of Georgi \& Politzer, 
$A_n^{{\rm (SM')}}\to A_n$, with the unphysical region contributing
in the calculation of the physical structure functions.  Hence, we
arrive at a conundrum: imposing the correct threshold leads to a
$Q^2$ dependent $A_n$, in contradiction to the standard OPE;
retaining the partonic interpretation of $A_n$ gives rise, on the
other hand, to unphysical contributions to the physical structure
functions, which one hopes will be canceled by higher twists.

%---------------------------------------------
\begin{figure}[!t]
\begin{center}
\vspace{40pt}
\includegraphics[scale=0.50,angle=270]{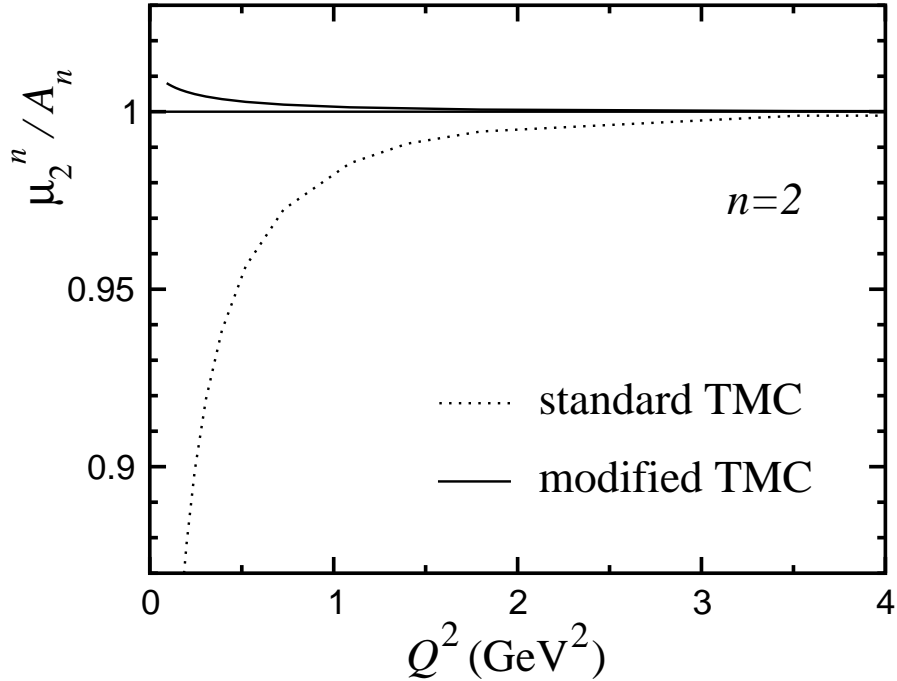}
\caption{
 Ratio of the $n=2$ Nachtmann moment of the $F_2$ structure
 function and the $n=2$ moment of the quark distribution,
 as a function of $Q^2$.  The curves correspond to standard TMC
 prescription \cite{Georgi:1976ve} (dotted) and the modified TMC
 (SM) prescription from Ref.~\cite{Steffens:2006ds}.
}
\label{fig:F2TMmom}
\end{center} 
\end{figure}
%---------------------------------------------

The large-$x$ behavior of the distributions is also particularly
important for the comparison of the Nachtmann moments $\mu_i^n$
of structure functions with the moments of the parton distributions.
The Nachtmann moments are, by construction, meant to be protected
from target mass effects, so that these should be equal to the
moments of the scaling parton distributions for {\it any} $Q^2$.
A comparison of the Nachtmann moment with $A_n$ was already
shown in Fig. \ref{fig:cnNacht}, and in Ref.~\cite{Steffens:2006ds}
this property was investigated in several other scenarios including
the modified definition of the parton distribution in Eq.\ (\ref{eq:TD}).

In the standard approach, Eq.\ (\ref{eq:GP}), there is an increasing
difference between the Nachtmann moments and the moments of
the parton distributions for $Q^2 < 1$~GeV$^2$, as can be seen in
the ratio shown in Fig.~\ref{fig:F2TMmom}.  At $Q^2=1$~GeV$^2$,
the difference is $\approx 2\%$.  In the modified approach, on the
other hand (Eq.~(\ref{eq:TD})), the Nachtmann moments are
essentially equivalent to the $\xi_0$-dependent
$A_n^{{\rm (SM)}}$ for $Q^2$ as low as 0.5~GeV$^2$, as shown
with the solid line.

The deviation of the ``standard TMC'' ratio from unity below
$Q^2 \sim 2-3$~GeV$^2$ suggests that the equality between the
Cornwall--Norton moments of the PDFs (or massless limit structure
functions) and the Nachtmann moments of the target mass corrected
structure functions will only hold at sufficiently large $Q^2$.
For $Q^2\gsim 2$~GeV$^2$, we can expect uncertainties due to
the treatment of TMCs to be $\lesssim 1\%$. These effects will also
appear when the evaluations are done using experimental data.
For lower values of $Q^2$, the direct connection between
Nachtmann moments of $F_2^{\rm TMC}$ and the parton model 
requires a more sophisticated treatment of the large-$x$ regime.

%%%%%%%%%%%%%%%%%%%%%%%%%%%%%%%%%%%%%%%%%%%%%%%%%%%%%%%%%%%%%%%%%%%
%%%%%%%%%%%%%%%%%%%%%%%%%%%%%%%%%%%%%%%%%%%%%%%%%%%%%%%%%%%%%%%%%%%
%%%%%%%%%%%%%%%%%%%%%%%%%%%%%%%%%%%%%%%%%%%%%%%%%%%%%%%%%%%%%%%%%%%

\section{Quantitative Effects of Target Mass Corrections \label{sec:results}}

The target mass corrections described in the previous sections, and
summarized in the master equations (\ref{eq:master}), are applied
in this section to neutrino--nucleon and electron--proton scattering.
The theoretical evaluation of the target mass corrected structure
functions $F_2$ and $F_3$ are compared with data from the NuTeV
neutrino--iron scattering experiment \cite{Tzanov:2005kr}.  (In this
discussion we neglect nuclear effects on structure functions, and
assume equality between the nuclear and nucleon structure functions
--- for a discussion of nuclear effects in 
inclusive lepton--nuclear scattering, see {\em e.g.}
Ref.~\cite{Kulagin:2004ie}.)  We also show results for the
electromagnetic structure function $F_2^p$ in electron--proton
scattering, with a comparison data.  We begin in
Sec.\ \ref{sec:results1}. with the ratio of $F_2$ and $F_3$ evaluated in $\nu N$
scattering ($N=(p+n)/2$) with and without target mass corrections.

%%%%%%%%%%%%%%%%%%%%%%%%%%%%%%%%%%%%%%%%%%%%%%%%%%%%%%%%
\subsection{TMC Effects in the Massless Quark Limit}
\label{sec:results1}

%---------------------------------------------
\begin{figure}[!t]
\begin{center}
\includegraphics[scale=0.75]{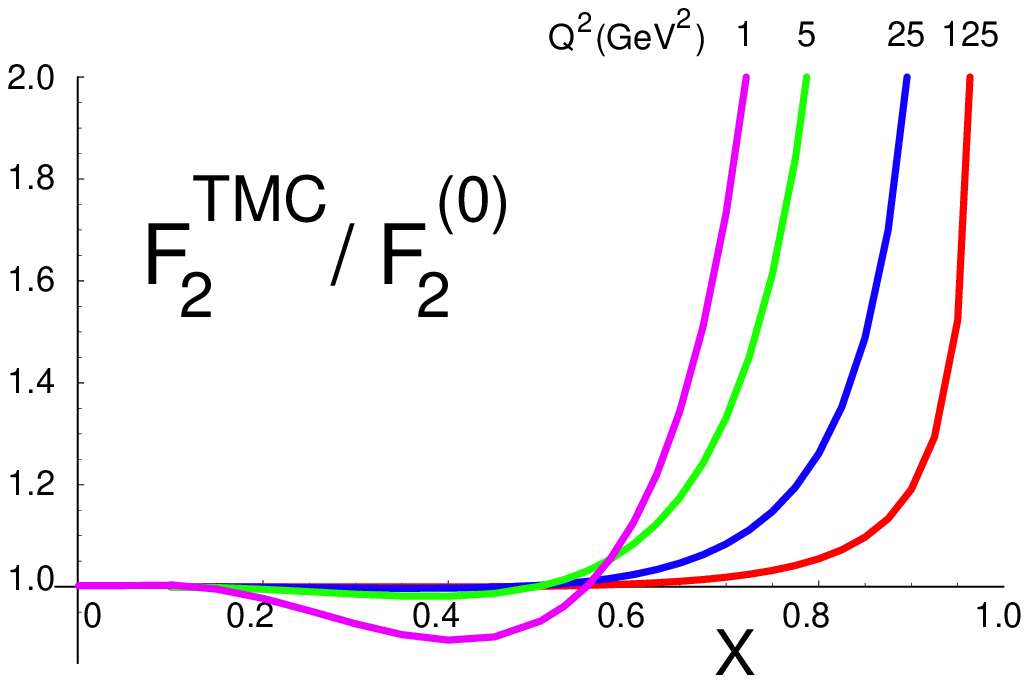}
\\[50pt]
\includegraphics[scale=0.75]{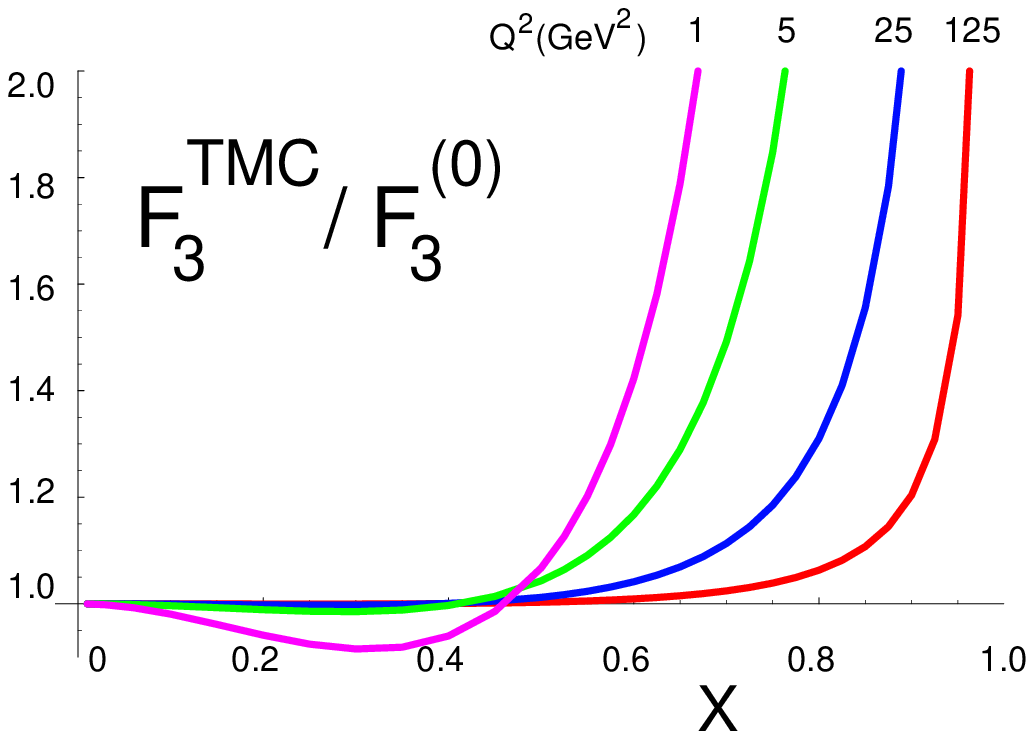}
\caption{Ratio of the $F_{2}$  and $F_{3}$  structure functions
	with and without target mass corrections
	($F_i^{\TMC}/F_i^{(0)},\ i=2,3$)
 vs.\ $x$ for $Q^2=\{1,5,25,125\}~\GeV^2$.
}
\label{fig:F2TM}
\end{center}
\end{figure}
%---------------------------------------------

Target mass corrections are relevant especially in the high-$x$ and
low-$Q^2$ regions. To quantify the size of the corrections, we evaluate
the structure functions at leading order in QCD, for neutrino--nucleon
scattering in the massless quark limit. For convenience we use the
CTEQ5 PDF set \cite{Lai:1999wy}; using other parameterizations has
a minimal effect on our numerical results.
The effects of the target mass corrections are illustrated in Fig.~\ref{fig:F2TM},
where the ratios of the $F_2$ and $F_3$ structure functions with and
without TMC effects are shown, namely $F_i^{\rm TMC} / F_i^{(0)},\ i=2,3$,
for $Q^2 = 1, 5, 25$ and $125~\GeV^2$.

The ratios of structure functions with TMCs to those without rise above
unity at large $x$, with the rise beginning at larger values of $x$ as
$Q^2$ increases. The correction can be quite large: for $x=0.8$,
for example, the TMC effect is $\sim30\%$ at $Q^2=5~\GeV^2$.  A large
part of the correction comes from the shift of $x\to \xi< x$ in the PDFs.
The PDFs are rapidly falling functions at large $x$, so even a small
change in the argument of the PDFs ($x$ replaced by $\xi$) can
have a significant impact on $F_i^{\rm TMC}/F_i^{(0)}$.
For $Q^2=1~\GeV^2$, $\xi$ deviates from $x$ even for $x\sim 0.3$
(see Fig.~\ref{fig:xi}).  
In the following section we discuss the relative importance of the
non-leading terms in the master equations (\ref{eq:master}).

%%%%%%%%%%%%%%%%%%%%%%%%%%%%%%%%%%%%%%%%%%%%%%%%%%%%%%%%
\subsection{Non-leading Terms in the Master Equations}

The non-leading (second and third) terms in Eqs.~(\ref{eq:f1})--(\ref{eq:f3})
constitute a small correction to the leading term. Since the evaluation of the
convolution integrals is quite time-consuming, it is useful to have an upper
bound for the size of these terms.  We can rewrite Eq.~(\ref{eq:f2}) as:
\begin{eqnarray*}
F_{2}^{{\rm TMC}}(x,Q^{2}) & = & \frac{x^{2}}{\xi^{2}r^{3}}F_{2}^{(0)}(\xi)
\bigg[1+\frac{6\mu x}{r}\ \frac{\xi^{2}h_{2}(\xi)}{F_{2}^{(0)}(\xi)}+
\frac{12\mu^{2}x^{2}}{r^{2}}\ \frac{\xi^{2}g_{2}(\xi)}{F_{2}^{(0)}(\xi)}
\bigg]
\end{eqnarray*}
with $\mu=M^{2}/Q^{2}$. 
The structure function $F_2^{(0)}$ appearing in the integrals in
Eqs.~(\ref{eq:defh2}) and (\ref{eq:defg2}) is a decreasing function
of $x$ or $\xi$ (see {\em e.g.} Fig.~\ref{fig:f2fit1} below).
Consequently, $F_2^{(0)}$ can be evaluated at the lower
integral limit, giving
$h_{2}(\xi) < (F_{2}^{(0)}(\xi) / \xi) (1-\xi)$ and
$g_{2}(\xi) < F_{2}^{(0)}(\xi)(-\ln \xi - 1 +\xi)$.
One then arrives at the following inequality:
\begin{eqnarray}
F_{2}^{{\rm TMC}}(x,Q^{2})
&<&
\frac{x^{2}}{\xi^{2}r^{3}}F_{2}^{(0)}(\xi)
\bigg[1+\frac{6\mu x \xi}{r}(1-\xi)+\frac{12\mu^{2}x^{2} \xi^2}{r^{2}}(-\ln \xi - 1+\xi)\bigg]\, .
\nonumber\\
\label{eq:f2bound}
\end{eqnarray}
The expressions $(6\mu x \xi/r) (1-\xi)$ and
$(12\mu^{2}x^{2}\xi^2 / r^{2})(-\ln \xi -1+\xi)$ can be easily evaluated
to obtain an upper bound for the contribution of the non-leading terms. 
Following the same line of argumentation one finds for the structure
function $F_3$:
\begin{equation}
F_{3}^{{\rm TMC}}(x,Q^{2}) < \frac{x}{\xi r^{2}}F_{3}^{(0)}(\xi)
\bigg[1-\frac{2\mu x \xi}{r}\ln \xi\bigg]\ .
\label{eq:f3bound}
\end{equation}
While the upper bounds for $F_2^{\rm TMC}$ and $F_3^{\rm TMC}$
are strictly satisfied for $x$ and $Q^2$ values relevant for target mass
corrections, these bounds are of limited practical use. 
For example, for $Q^2=1~\GeV^2$, Eq.~(\ref{eq:f2bound}) places
a limit on the non-leading corrections to be less than $\sim 65\%$
of the leading term at large $x$. The actual value is much less,
below $\sim 21\%$.  Therefore it is useful to note that 
\begin{equation}
F_{2}^{{\rm TMC}}(x,Q^{2}) \simeq \frac{x^{2}}{\xi^{2}r^{3}}F_{2}^{(0)}(\xi)
\bigg[1+\frac{6\mu x \xi}{r}(1-\xi)^2 \bigg]
\label{eq:f2tmc_approx}
\end{equation}
provides a very good approximation of the structure function
$F_2^{\rm TMC}(x,Q^2)$.
Similarly, $F_3^{\rm TMC}(x,Q^2)$ can be approximated by
\begin{equation}
F_{3}^{{\rm TMC}}(x,Q^{2}) \simeq \frac{x}{\xi r^{2}} F_{3}^{(0)}(\xi)
\bigg[1-\frac{\mu x \xi}{r} (1-\xi) \ln \xi \bigg]\, .
\label{eq:f3tmc_approx}
\end{equation}

%%%%%%%%%%%%%%%%%%%%%%%
% Discussion of Figures
%%%%%%%%%%%%%%%%%%%%%%%
%---------------------------------------------
\begin{figure}[!t]
\begin{center}
\includegraphics[scale=0.5]{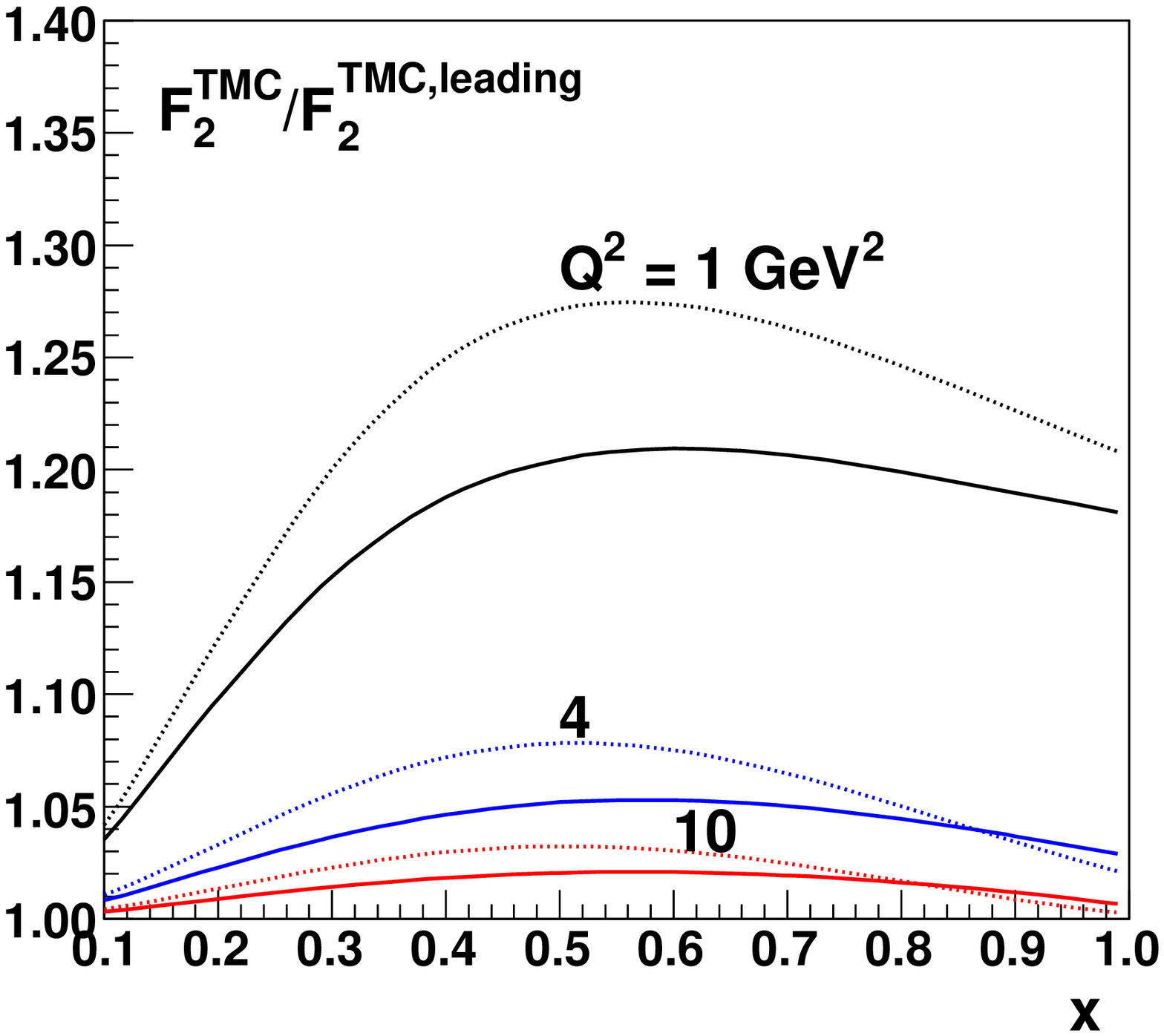}
\includegraphics[scale=0.5]{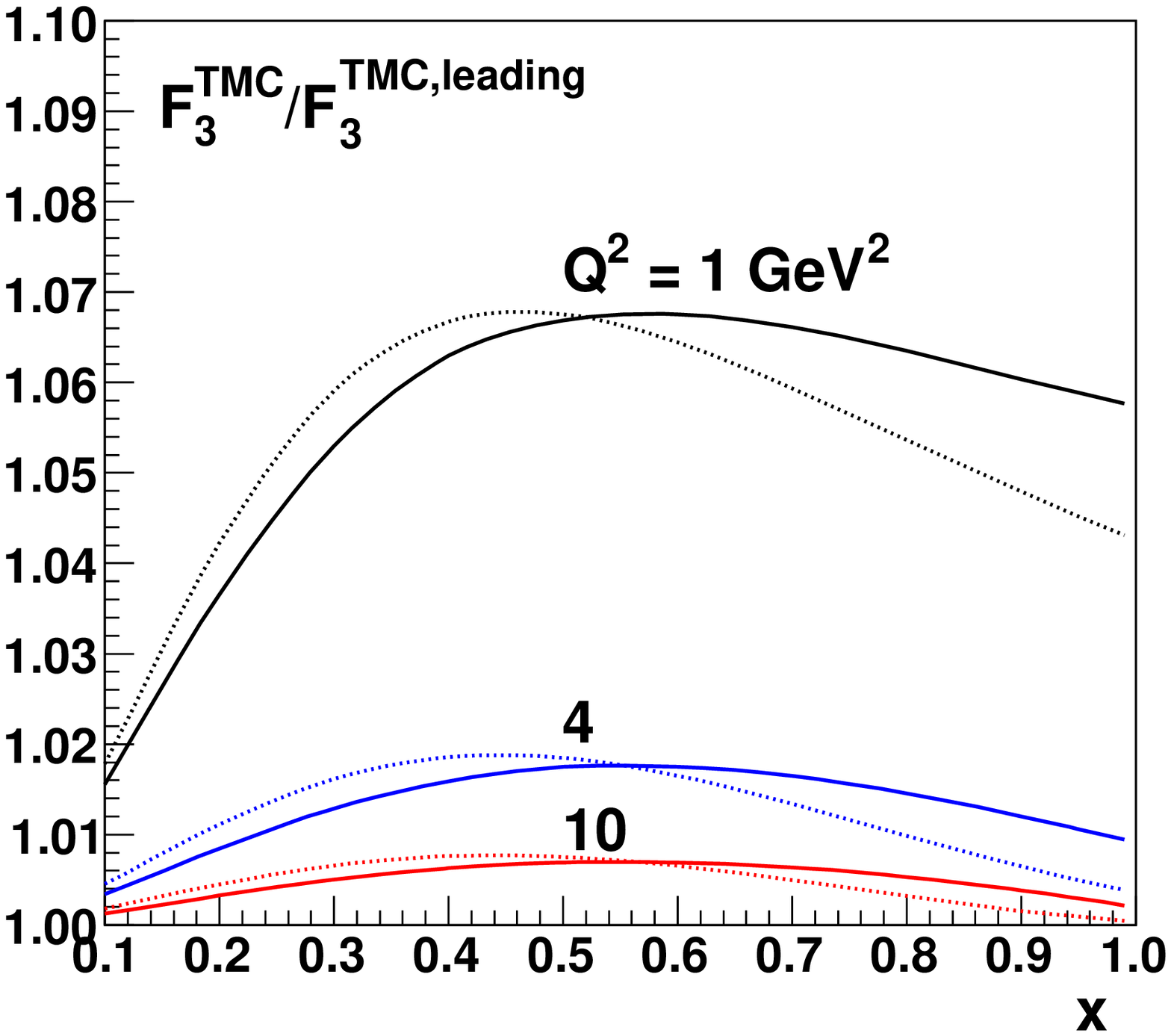}
%\vspace*{-1cm}
\caption{Ratio of the target mass corrected $F_2^{\rm TMC}(x,Q^2)$ (top)
	and $F_3^{\rm TMC}(x,Q^2)$ (bottom) structure functions to the leading
	contributions in Eqs.~(\ref{eq:f2}) and~(\ref{eq:f3}), at $Q^2 = 1, 4$ and
	10~GeV$^2$.  The solid curves represent the exact results, while the
	dotted curves have been calculated using the approximate formulas
	in Eqs.~(\ref{eq:f2tmc_approx}) and~(\ref{eq:f3tmc_approx}).
}
\label{fig:tmcsize} 
\end{center}
\end{figure}
%---------------------------------------------

The magnitude of the non-leading contributions to the target mass
correction is illustrated in Fig.~\ref{fig:tmcsize}, where the ratio of the
target mass corrected $F_2$ (top graph) and $F_3$ (bottom graph)
structure functions is shown relative to the leading contribution, at
$Q^2 = 1, 4$ and 10~GeV$^2$.  Here
$F_2^{\rm TMC, leading}(x,Q^2) = (x^2 / \xi^2 r^3) F_2^{(0)}(\xi)$
represents the leading contribution to the target mass corrected
structure function $F_2^{\rm TMC}(x,Q^2)$ ({\em cf.} Eq.~(\ref{eq:f2})),
while the corresponding results for the $F_3$ structure function is
given by
$F_3^{\rm TMC, leading}(x,Q^2)= (x / \xi r^2) F_3^{(0)}(\xi)$.
For definiteness, the structure functions are for charged current
neutrino--proton scattering and have been computed in next-to-leading
order of QCD including quark mass effects.  However, the results are
very robust concerning variations of these details (process, order,
quark mass effects).
The dotted curves in Fig.~\ref{fig:tmcsize} present the results of
the approximate formulas in Eqs.~(\ref{eq:f2tmc_approx})
and~(\ref{eq:f3tmc_approx}).  As can be seen, the simple
approximations are in very good agreement with the exact results.

The excess over unity depicts the fractional contribution of the
non-leading terms.  Clearly, the non-leading contributions to the
structure function $F_2^{\rm TMC}$ are relatively small and positive.
For $Q^2=1~\GeV^2$ they amount about $21\%$.  However, for
$Q^2=4~\GeV^2$ they correct the leading term already by less
than $6\%$, and for $Q^2=10~\GeV^2$ by less than $3\%$.
The major contribution from the non-leading pieces comes from
the terms proportional to $h_2$ or $h_3$, whereas the part
proportional to $g_2$ constitutes a small correction.
These results imply that $F_2^{\rm TMC}(x,Q^2)$ can be approximated
in many cases by $F_2^{\rm TMC, leading}(x,Q^2)$.  Moreover, if more
precision is needed, the non-leading pieces can be simulated, to good
accuracy, by the simple approximation in Eq.~(\ref{eq:f2tmc_approx}).
For $F_3^{\rm TMC}$, the non-leading terms are even smaller,
contributing less than $7\%$ for $Q^2=1~\GeV^2$. For $Q^2=4~\GeV^2$
they are already below the $2\%$ level.

The findings of this section are useful in providing a simple
explanation of the main effects of the target mass corrections
shown in Fig.~\ref{fig:F2TM}.
In fact, to good accuracy, the ratio in Fig.~\ref{fig:F2TM} is given by:
\begin{equation}
{F_2^{\rm TMC}(x) \over F_2^{(0)}(x)}
\simeq \frac{x^2}{\xi^2 r^3} {F_2^{(0)}(\xi) \over F_2^{(0)}(x)}\, .
\label{eq:explain}
\end{equation}
Since $\xi < x$, and $F_2$ is a monotonically decreasing function,
one can see that the simple shift from $x$ to $\xi$ in the argument
of $F_2^{(0)}$ (along with the factor $x^2/\xi^2 > 1$) produces the
enhancement over unity visible in this figure.  The dip below unity
for low $Q^2$ values and intermediate $x$ can be explained by
the factor $1/r^3$ in Eq.~(\ref{eq:explain}) which is smaller than one.

%%%%%%%%%%%%%%%%%%%%%%%%%%%%%%%%%%%%%%%%%%%%%%%%%%%%%%%%
\subsection{TMC Effects in NuTeV Structure Functions}

%---------------------------------------------
\begin{figure}[!t]
\begin{center}
\includegraphics[scale=0.80]{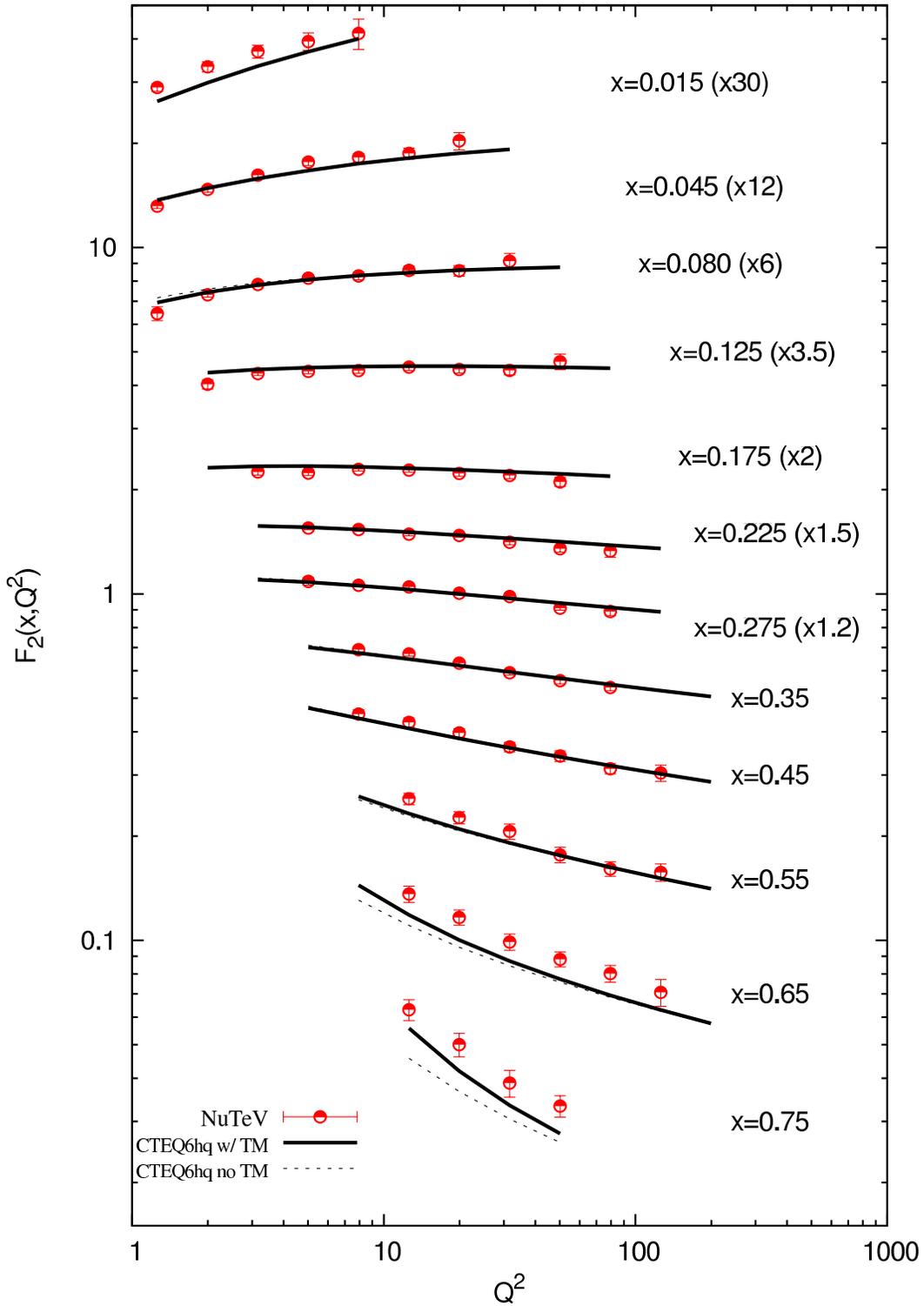}
\caption{Comparison of the $F_2$ structure function, with and without
	target mass corrections, and NuTeV data \cite{Tzanov:2005kr}.
	The base PDF set is CTEQ6HQ \cite{Kretzer:2003it}.
}
\label{fig:f2nutev} 
\end{center}
\end{figure}
%-----------------

A comparison of the $F_2$ structure function with the NuTeV neutrino
data \cite{Tzanov:2005kr} is shown in Fig.~\ref{fig:f2nutev}, where $F_2$
is calculated both with (solid curves) and without (dashed curves) target
mass corrections. For this comparison, the full NLO evaluation is made
including charm quark mass effects.  The PDF set for this comparison was
taken to be CTEQ6HQ \cite{Kretzer:2003it}.  As Fig.~\ref{fig:f2nutev} shows,
there is a clear improvement in the agreement in the high-$x$ region
between the NuTeV data and NLO predictions including target mass effects.

%%%%%%%%%%%%%%%%%%%%%%%%%%%%%%%%%%%%%%%%%%%%%%%%%%%%%
%%%%%%%%%%%%%%%%%%%%%%%%%%%%%%%%%%%%%%%%%%%%%%%%%%%%%
\subsection{Unfolding Target Mass Effects From Structure Function Data}

%---------------------------------------------
\begin{figure}[!t]
\begin{center}
\includegraphics[scale=0.70]{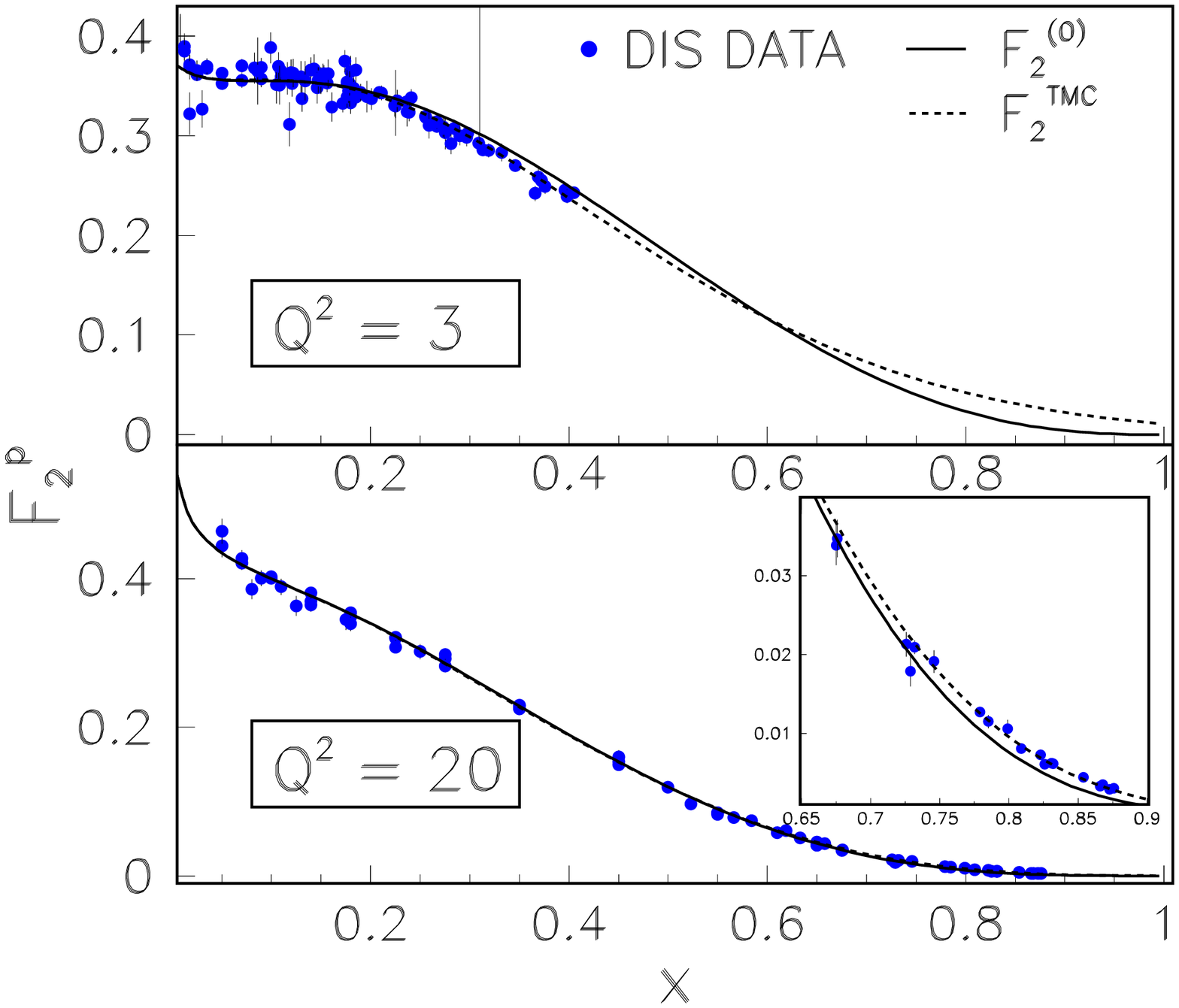}
\caption{Results of the unfolding procedure of Ref.~\cite{christy:TM}
	for the proton $F_2^p$ structure function, at $Q^2=3~\GeV^2$
	(top panel) and $Q^2=20~\GeV^2$ (bottom panel).  The massless
	limit functions are shown by the solid curves, and the target mass
	corrected results by the dashed curves.
}
\label{fig:f2fit1}
\end{center}
\end{figure}
%---------------------------------------------

%---------------------------------------------
\begin{figure}[!t]
\begin{center}
\includegraphics[scale=0.6]{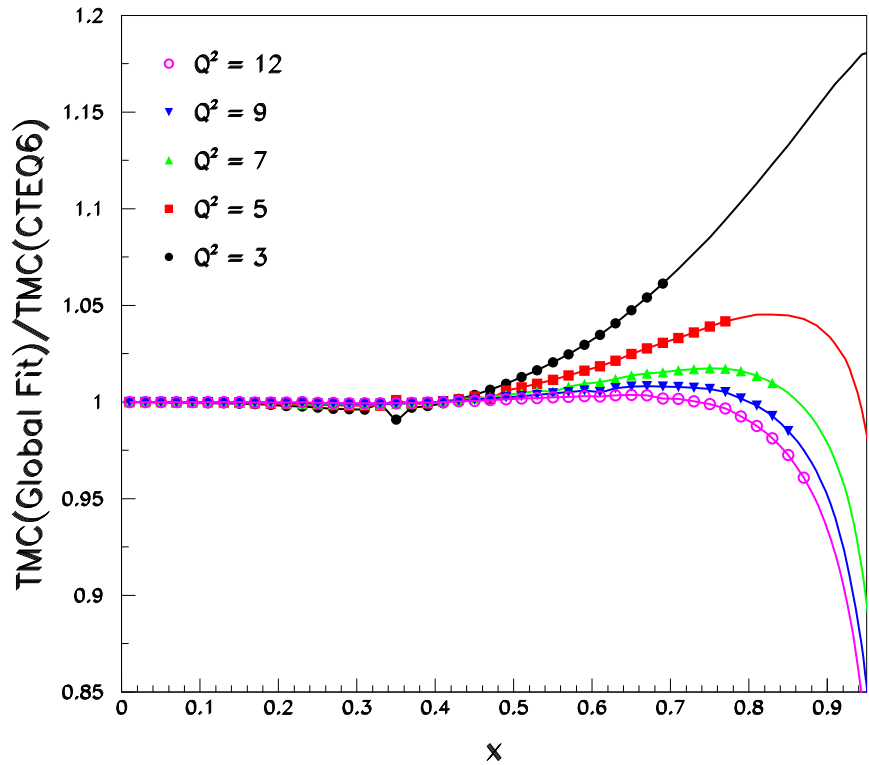}
\caption{Ratio of target mass corrections for $F_2^p$ determined from the
	unfolding procedure (global fit) of Ref.~\cite{christy:TM} to the corrections
	obtained from CTEQ6 PDFs {\it cf.} Eq.~(\protect\ref{eq:doubleratio}).
	The $Q^2$ values shown are in GeV$^2$.
	The $x$ region for which $W^2~>~4~\GeV^2$ is indicated by the
	symbols. Note that even at high $x$ and low $Q^2$ the difference
	in the TMCs is typically less than several percent.}
\label{fig:christy_cteq_ratio}
\end{center}
\end{figure}
%---------------------------------------------

%---------------------------------------------
\begin{figure}[!t]
\begin{center}
\includegraphics[scale=0.6]{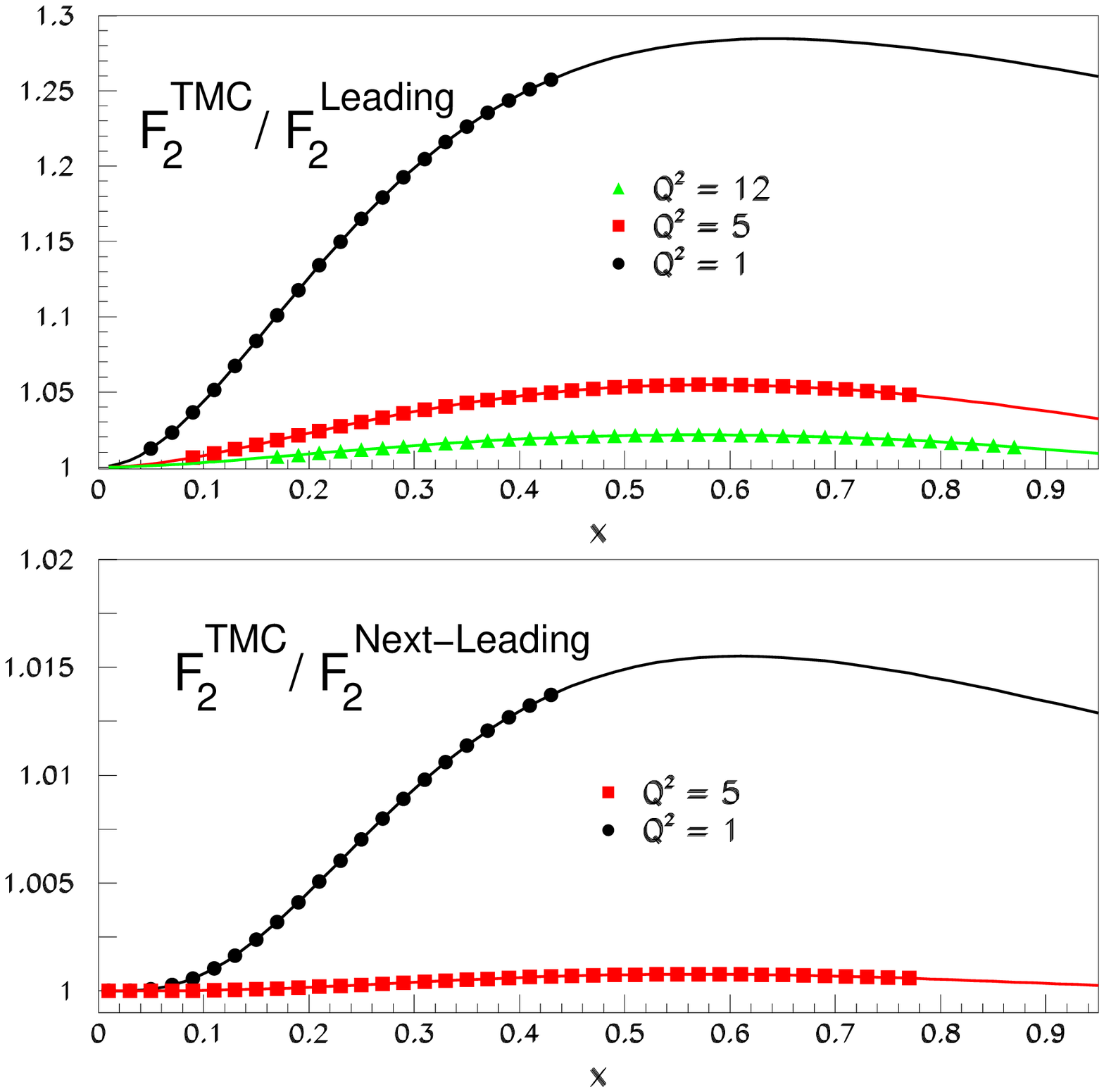}
\caption{Ratio of the full $F_2^{\rm TMC}$, as extracted from the unfolding procedure,
	to that including only the leading term $F_2^{\rm TMC,Leading}$ (top panel), 
         and next-to-leading term $F_2^{\rm TMC,Next-Leading}$
	(bottom panel), for $Q^2=\{1, 5, 12 \}~\GeV^2$.  The $x$ region for which
	$W^2~>~4~\GeV^2$ is indicated by the symbols.  Note the change of scale
	between the two panels.
}
\label{fig:nlt_f2fit} 
\end{center}
\end{figure}
%---------------------------------------------

In neutrino and charged lepton scattering from nucleons it is the
$F_i^{\rm TMC}(x,Q^2)$ functions that are measured by experiment.
In principle, measurements over a range in $x$ at fixed $Q^2$ allow for
a determination of the massless limit structure functions $F_i^{(0)}(x)$ by
inverting the master equations of Sec.\ \ref{sec:OPE}.  Implicit in such a procedure is
the assumption that dynamical higher twists operators either contribute
negligibly to the data in the region of interest, or that they also obey the
master equations for the TMCs.  Under such conditions the unfolding
procedure then allows both the $F_i^{(0)}(x)$ and the target mass
contributions to be determined directly from data.  The procedure is to
parameterize $F_i^{(0)}$(x) at fixed $Q^2$, insert this into the master
equations, and then minimize the difference of the resulting
$F_i^{\rm TMC}$ with respect to the data.

This method is complementary to PDF fits.  One might expect the
$F_2^{(0)}$ structure function so extracted to be consistent with that
obtained from PDF fits to the data when including TMCs.  However,
an advantage of the unfolding procedure is that the separation of
the $F_2^{(0)}$ (which is described by pQCD) from the target mass
contributions is good to all orders in $\alpha_s$, since Eq.~(\ref{eq:f2})
is valid to all orders.  Comparisons of the results from the two
techniques could help resolve any possible mixing between
contributions from TMCs, higher twists, and those from higher order
pQCD terms.  Such an unfolding procedure has been undertaken 
\cite{christy:TM} for the world data set
of charged lepton scattering data from the proton, and the results
are currently being prepared for publication.

In this study the $F_2$ data were fitted globally for
$0.5~<~Q^2~<~250~\GeV^2$ by allowing for a $Q^2$ dependence of
the parameters describing $F_2^{(0)}$(x). The results for $Q^2 = 3$
(top panel) and $20~\GeV^2$ (bottom panel) are shown in
Fig.~\ref{fig:f2fit1}.  The solid curve is the $F_2^{(0)}$ determine
from the fit, while the dashed curve is the full $F_2^{\rm TMC}$.
Consistent with the determination from PDF fits previously discussed,
the TMC contributions to $F_2$ are large at small $Q^2$, as much as
9\% even at $x = 0.4$ for $Q^2 = 3~\GeV^2$.  While the TMCs become
much smaller at higher $Q^2$, they are still sizable at higher $x$,
as can be seen in the inset in Fig.~\ref{fig:f2fit1} (bottom panel).
At $Q^2 = 20~\GeV^2$ the contributions from TMCs are 4\%, 8\%
and 14\% at $x$ = 0.65, 0.70 and 0.75, respectively.

It is interesting to note that even in kinematic regions where the TMCs
are large, the unfolding procedure gives results for the target mass
contributions which are quite consistent with that determined by
inserting existing CTEQ6 PDFs into the master equation for $F_2$.
This can be observed in Fig.~\ref{fig:christy_cteq_ratio}, where the
ratio of TMC factors determined from the unfolding procedure to
those calculated from CTEQ6 PDFs are plotted.  The correction
factors are here defined as $F_2^{(0)} \, / \, F_2^{\rm TMC}$,
{\it e.g.} the multiplicative factor which would be applied to data to
yield a measured $F_2^{(0)}$.  Figure~\ref{fig:christy_cteq_ratio}
shows the double ratio 
\begin{equation}
\frac{
\left( F_2^{(0)}\, / \, F_2^{\rm TMC} \right)_{\rm unfold}
}{
\left( F_2^{(0)}\, / \, F_2^{\rm TMC}\right)_{\rm CTEQ6}
}
\label{eq:doubleratio}
\end{equation}
as a function of $x$ for $3~<~Q^2~<~9~\GeV^2$.  For the region
$W^2~>~4~\GeV^2$ (indicated by the symbols) the difference is
typically less than a few percent even up to $x~\approx~0.9$.

As a final note, the size of the non-leading terms from the unfolding
procedure is shown in Fig.~\ref{fig:nlt_f2fit} as a function of $x$ for
$Q^2$ in the range 1 to 12~$\GeV^2$.  We display the ratio of the
full $F_2^{\rm TMC}$ to that including only the leading term (top panel),
and next-to-leading term (bottom panel); note the change of scale
between the two panels.  The fractional contribution from the leading
term is found to be in excellent agreement to that shown in
Fig.~\ref{fig:tmcsize}, while the contribution from the double integral
term is at most 1.5\% at $Q^2=1~\GeV^2$, and neglible for higher $Q^2$.

%\clearpage
%%%%%%%%%%%%%%%%%%%%%%%%%%%%%%%%%%%%%%%%%%%%%%%%%%%%%

%%%%%%%%%%%%%%%%%%%%%%%%%%%%%%%%%%%%%%%%%%%%%%%%%%%%%
\subsection{Longitudinal Structure Function} 

%%%%%%%%%%%%%%%%%%%%%%%
% Discussion of Figures
%%%%%%%%%%%%%%%%%%%%%%%
%---------------------------------------------
\begin{figure}[!t]
\begin{center}
\includegraphics[scale=0.5]{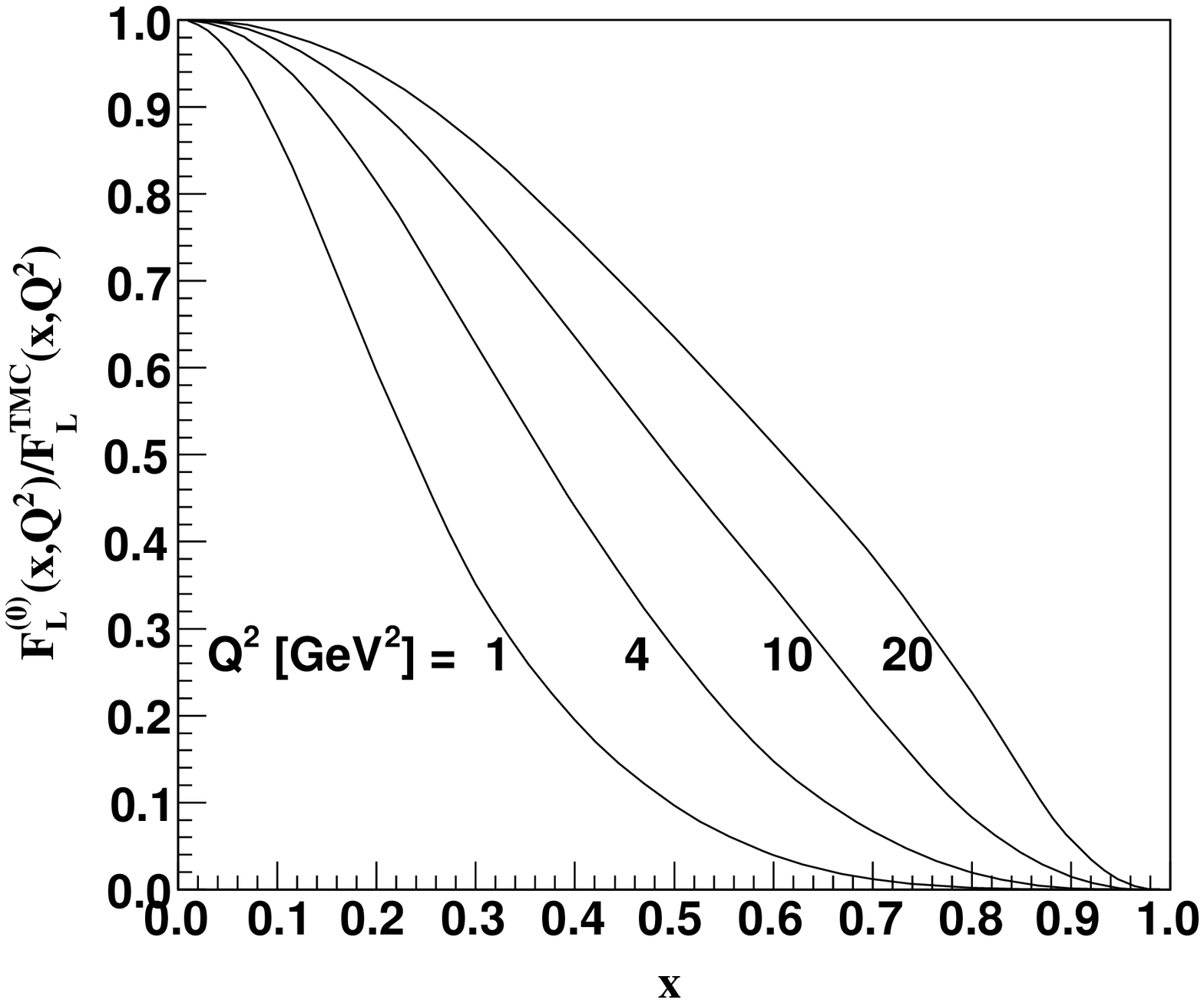}

\vspace{-1cm}
\includegraphics[scale=0.5]{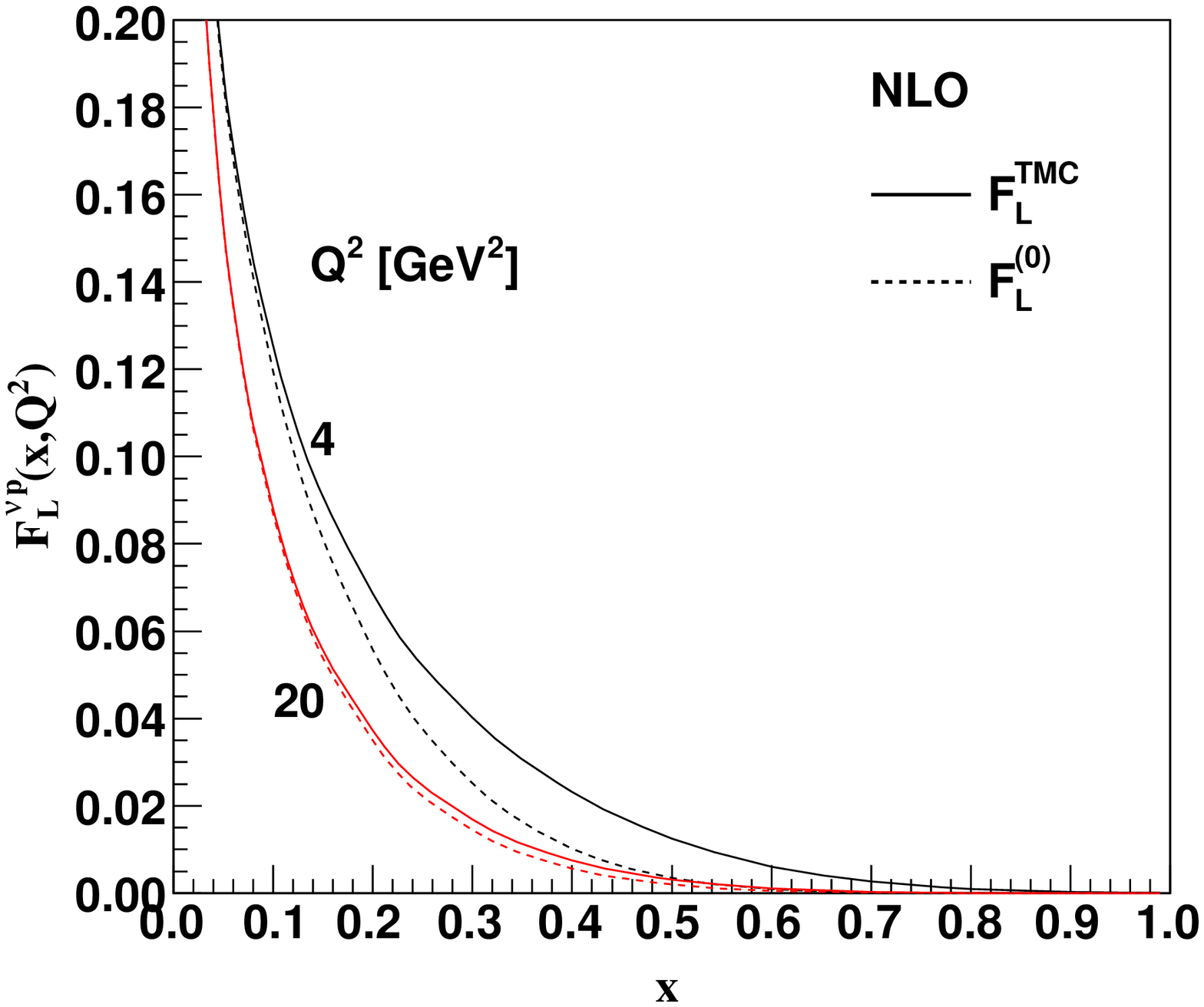}
%\vspace*{-1cm}
\caption{
a) Ratio of the structure function $F_L$ without ($F_L^{(0)}(x,Q^2)$) and
with ($F_L^{\rm TMC}(x,Q^2)$) target mass corrections at $Q^2 = 1, 4, 10,$ and
	20~GeV$^2$ vs.\ $x$  
\qquad
b) Absolute magnitude of the target mass corrected $F_L^{\rm TMC}(x,Q^2)$ (solid) 
as compared with the structure function 
in the limit of a vanishing target mass $F_L^{(0)}(x,Q^2)$ (dashed) at $Q^2 = 4$ and 20~GeV$^2$
vs.\ $x$.
}
\label{fig:flong} 
\end{center}
\end{figure}
%---------------------------------------------

Finally, we show a set of plots for the longitudinal structure function, $F_L$, in 
Fig.~\ref{fig:flong}.  
The target mass corrected structure function 
$F_L^{\rm TMC}(x,Q^2)$ has been computed according
to Eq.\ (\ref{eq:fl}) using parton model results for the structure function 
$F_L^{(0)}(x,Q^2)$ for charged current neutrino--proton scattering in NLO QCD
including quark mass effects.
Because the massless leading-order parton model contributions to $F_L^{(0)}$ vanish, this 
quantity is ideally suited to study both mass effects and higher order corrections. 

The effects of the TMC are illustrated in Figure \ref{fig:flong}a) where the 
ratio $F_L^{(0)}/F_L^{\rm TMC}$ is shown for $Q^2=1,4,10$, and 20~GeV$^2$.
Note that the present ratio is the inverse of similar ratios shown in 
Fig.\ \ref{fig:F2TM} for the structure functions $F_2$ and $F_{3}$.
As can be seen, the curves are steeply falling, implying very large target mass
corrections.
%
%However, this figure portrays an unrealistic situation at large $x$; ...
%
While the TMC are 
proportionately large in this region, unfortunately the longitudinal structure function
becomes vanishingly small. This point is evident in  Fig.~\ref{fig:flong}b) where 
we plot the absolute magnitude of $F_L$. 
Here, the solid lines represent the full target mass corrected $F_L^{\rm TMC}(x,Q^2)$, 
and  the dashed lines represents the conventional parton model $F_L^{(0)}(x,Q^2)$.
We do observe for $Q^2=4~\GeV^2$ that there are significant differences 
in the intermediate $x$ region ($\sim 0.3$); however, these effects are negligible 
for  $Q^2=20~\GeV^2$.
In conclusion, we note that while the longitudinal structure function,
$F_L$, is difficult to measure because of its small size, a precise
measurement of this quantity could prove to be sensitive to the TMC
and quark mass effects.\cite{Liang:2004tj}

%%%%%%%%%%%%%%%%%%%%%%%%%%%%%%%%%%%%%%%%%%%%%%%%%%%%%%%%%%%%%%%%%%%
%%%%%%%%%%%%%%%%%%%%%%%%%%%%%%%%%%%%%%%%%%%%%%%%%%%%%%%%%%%%%%%%%%%
%\input{7conclusions}

\section{Conclusions}
\label{sec:conclusions}

We have presented in this review a survey of the key issues pertaining to
target mass corrections (TMCs) in 
inclusive lepton--nuclear scattering 
structure functions and
their impact on the analysis of experimental results.  As illustrated by our
structure function results with and without TMC terms
(Figs.~\ref{fig:f2nutev}--\ref{fig:nlt_f2fit}), current experimental accuracy
demands that these effects be properly incorporated.

In the context of the operator product expansion (OPE), we have outlined
the derivation of the master equation for the target mass corrected
structure functions $F_j^{\rm TMC}(x,Q^2)$, Eq.~(\ref{eq:master}),
without any assumptions about fixed-order perturbation theory or the
Callan--Gross relation.
However, when relating the OPE results to calculable parton model
expressions, we are forced to make compromises; for example, we
generally work with only at leading twist ({\it cf.} Sec.~\ref{sec:OPEpm})
and at a finite order of $\alpha_s$ in perturbation theory.
Because the TMCs enter as $(M^2/Q^2)^j$ corrections, it is essential to
organize our expansion so that these terms can be distinguished from
true higher twist contributions, such as from four-quark operators.

To illustrate the dependence on both hadron and parton masses in the
formalism, we present explicit results for heavy quark production
via neutrino--nucleon scattering.
For the non-leading terms, we compute an upper limit for the integral
terms, derive approximate expressions, and evaluate the full expression
numerically; as expected, the dominant TMC effects arise at large $x$
and small $Q^2$ values.

In the limit $x\to1$ we encounter an additional complication in the
standard TMC formulation, associated with the non-vanishing of
the target mass corrected structure function at $x=1$.
This problem arises if the massless limit structure function,
$F_2^{(0)}(\xi,Q^2)$, is not required to vanish in the kinematically
forbidden region $\xi > \xi_0$.  One may hope that higher twist terms
could cancel the unphysical contributions.  Alternatively, one can
enforce the constraint that $F_2^{(0)}(\xi,Q^2)$ does vanish for
$\xi > \xi_0$, which removes the unphysical contributions, but at the
expense of introducing additional $\xi_0$ dependence into the structure
function.  We have reviewed various attempts that have been made in
the literature to address this problem.

Finally, the importance of the TMCs for comparison with experimental
data is established in Sec.~\ref{sec:results}.
We perform fits of structure functions both with and without the TMC
contributions, and observe that these terms can have a substantive
impact in the region of high $x$.  While there are a number of
factors which enter these fits, it is encouraging to see that the TMC
contributions generally improve the agreement between the fits and data.

We also demonstrate (Figs.~\ref{fig:f2fit1}, \ref{fig:christy_cteq_ratio},
and~\ref{fig:nlt_f2fit}) quantitative agreement extracting the $F_2$
structure function with two complementary methods. We find that
the ratio of the TMCs for $F_2$ obtained starting with PDFs and including TMCs, {\it vs.}
starting with the TMC master equation, agrees at the 5\% or better
level, even for $Q^2 = 3~\GeV^2$ for $x<0.65$.

The fact that the experimental data are sufficiently accurate to be
sensitive to the TMC effects represents an important milestone in the
study of inelastic lepton--nucleon scattering. From this foundation,
future studies can more fully characterize the TMC effects, and begin
to separately identify higher twist contributions for these processes.

%%%%%%%%%%%%%%%%%%%%%%%%%%%%%%%%%%%%%%%%%%%%%%%%%%%%%%%%%%%%%%%%%%%
%%%%%%%%%%%%%%%%%%%%%%%%%%%%%%%%%%%%%%%%%%%%%%%%%%%%%%%%%%%%%%%%%%%
%%%%%%%%%%%%%%%%%%%%%%%%%%%%%%%%%%%%%%%%%%%%%%%%%%%%%%%%%%%%%%%%%%%
%------------------------
\section*{Acknowledgments}

We thank 
Stefan Kretzer, 
Jorge Morfin, 
Joseph (Jeff) Owens, 
A. Kataev,
Ingo Sick,
 for valuable discussions. 
%%%
F.I.O., I.S.,  and J.Y.Y.  acknowledge the hospitality of Argonne,
BNL, CERN, and Fermilab  where a portion of this work was performed.  
%%%
This work was partially supported by the U.S. Department of Energy
under grant 
DE-FG02-04ER41299, 
DE-FG02-91ER40685, 
DE-FG02-91ER40664, 
DE-FG03-95ER40908, contract
W-31-109-ENG-38; contract DE-AC05-06OR23177 (under which Jefferson
Science Associates LLC operates the Thomas Jefferson National
Accelerator Facility), the National Science Foundation grant 0400332,
the
Lightner-Sams Foundation, 
and
the Sam Taylor Foundation. 
%%%
The work of J.~Y.~Yu was supported by the
Deutsche Forschungsgemeinschaft (DFG) through grant No.~YU 118/1-1.

%------------------------

%%%%%%%%%%%%%%%%%%%%%%%%%%%%%%%%%%%%%%%%%%%%%%%%%%%%%%%%%%%%%%%%%%%
%%%%%%%%%%%%%%%%%%%%%%%%%%%%%%%%%%%%%%%%%%%%%%%%%%%%%%%%%%%%%%%%%%%
%%%%%%%%%%%%%%%%%%%%%%%%%%%%%%%%%%%%%%%%%%%%%%%%%%%%%%%%%%%%%%%%%%%
%\newpage{}
%\input{appendices.tex}

%%%%%%%%%%%%%%%%%%%%%%%%%%%%%%%%%%%%%%%%%%%%%%%%%
\appendix
\def\patch{\null}
\def\patch{\\}

\section{\patch Kinematics and Fundamental Relations\label{app:kin}}

%%%%%%%%%%%%%%%%%%%%%%%%%%%%%%%%%%%%%%%%%%%%%%%%%
\subsection{\patch Notation}
We consider the basic inclusive lepton--nuclear scattering 
 process $\ell(k) + N(P) \to \ell'(k') + X(P_X)$ 
({\it cf.}  Fig.~\ref{fig:DIS}), with $q=k-k'$ the four-momentum transferred
to the nucleon, and define $x$ to be the standard DIS scaling variable
in the Bjorken limit:
\begin{eqnarray}
x
&=&
\frac{-q^2}{2P\cdot q}
= \frac{Q^2}{2M\nu} \quad,
\label{eq:xbj}
\end{eqnarray}
where $-Q^2 = q^2$ is the virtuality of the exchanged boson, and $P$
is the target hadron four-momentum, with $M$ the target nucleon mass
($P^2=M^2$).  We denote by $m_i$ the mass of the $i$'th quark.  For a
nuclear target with atomic number $A$, it is convenient to rescale the
Bjorken variable $x$ by $A$ so that the denominator of Eq.~(\ref{eq:xbj})
is still the mass of a nucleon.  Additionally, we introduce the energy
transferred to the hadronic system in the target rest frame as:
\begin{eqnarray}
\nu
&=&
E-E'
= 
\frac{P\cdot q}{M}
\end{eqnarray}
and the inelasticity
\begin{eqnarray}
y
&=&
\frac{E-E'}{E}
=
\frac{P\cdot q}{P\cdot k}
\quad ,
\end{eqnarray}
where $E$ and $E'$ represent the initial and final lepton energies,
respectively.  The mass of the hadronic final state is given by:
\begin{eqnarray}
W^2
&=&
P_X^2
=
(P+q)^2
=
M^2 + 
\frac{Q^2}{x}(1-x)\ .
\end{eqnarray}
For convenience, we also introduce the variable $r$, which is a
combination of factors that appears frequently:
\[
r=\sqrt{1+\frac{4x^{2}M^{2}}{Q^{2}}}\equiv\sqrt{1+\frac{Q^{2}}{\nu^{2}}}
\quad .
\]

%%%%%%%%%%%%%%%%%%%%%%%%%%%%%%%%%%%%%%%%%%%%%%%%%
\subsection{\patch\noindent Generalized Nachtmann Variable}

 For massless quarks, the parton light-cone fraction is given by the
Nachtmann variable~$\xieta$.
Given the 4-vector $P_\mu=\{P_t,P_x,P_y,P_z\}$, we can re-express this
in light-cone coordinates as
$P_\mu=\{P_+,\overrightarrow{P}_\perp,P_-\}$ where\footnote{There are
multiple conventions here; sometimes the $1/\sqrt{2}$ is replaced by a
$1/2$ or a $1$.}  $P_\pm=(P_t\pm P_z)/\sqrt{2}$ and
$\overrightarrow{P}_\perp=\{P_x,P_y\}$ are the (boost-invariant)
transverse components.
Light-cone coordinates are a convenient representation when the
momentum components are strongly ordered, {\it e.g.} $P_+ \gg P_\perp,
P_-$.
Thus, if the hadron light-cone vector is
$P_\mu=\{P^+,\overrightarrow0,M^2/(2P^+)\}$, the parton vector for a
(massless) parton is $p_\mu=\{\xieta
P^+,\overrightarrow0,0\}$.\footnote{The parton model is derived in the
collinear limit where the parton transverse momentum is neglected,
{\it e.g.} $\overrightarrow{p}_\perp=0$; finite transverse momentum
can contribute to the TMCs,  
{\it cf.} Sec.~\ref{sec:intro}.
}
  In the case of massive partons, the Nachtmann variable $\xieta$ is
generalized to
$\bar\xieta$. \cite{Olness:1986mv,Aivazis:1993kh,Aivazis:1993pi}

We identify two generalized Nachtmann-type of variables, $\xi$ and $\bar{\xi}$,
which are related to the Bjorken $x$ variable via a dimensionless multiplicative
factor.  These can be written as:
\noindent \begin{eqnarray}
\xi & = & x\, R_M\ ,\\
\bar{\xi} & = & \xi\, R_{ij}=x\, R_{M}\, R_{ij}\ ,
\end{eqnarray}
\noindent where \noindent
\begin{eqnarray}
R_{M} & = & \frac{2}{1+\sqrt{1+(4x^{2}M^{2}/Q^{2})}}
	=\frac{2}{1+r}\ , \\
R_{ij} & = & \frac{Q^{2}-m_{i}^{2}+m_{j}^{2}
	+\Delta(-Q^{2},m_{i}^{2},m_{j}^{2})}{2Q^{2}}\ ,
\label{eq:Rij}\\
\Delta(a,b,c) & = & \sqrt{a^{2}+b^{2}+c^{2}-2(ab+bc+ca)}\ ,
\end{eqnarray}
where $m_i$ is the initial quark mass and 
$m_j$ is the final quark mass.
The variable $\xi=x\, R_{M}$ essentially corrects Bjorken $x$
for the effects of the hadronic mass. Computationally, this arises
from the final state momentum conserving delta-function,
$\delta^{4}(q+P-P_{X})$, which can be re-expressed to include
the delta-function $\delta(x-\xi/R_{M})$.

In a similar manner, the variable $\bar{\xi}=\xi\, R_{ij}$ further
corrects the $\xi$ variable for the effects of the partonic masses.
The origin of this correction is the momentum conserving
delta-function at the partonic level, $\delta^{4}(q+p_{i}-p_{j})$,
which can be re-expressed to include $\delta(\xi-\bar{\xi}/R_{ij})$.

It is important to note that the hadronic correction $R_{M}$ and
the partonic mass corrections $R_{ij}$ are separate and can be
applied multiplicatively.  Also note that the $R_{ij}$ correction
factor depends on the specific partonic masses involved; hence,
this cannot be simply factored out as with the $R_{M}$ terms.

In the limit the initial quark mass vanishes 
($m_{i}\to0$), the partonic correction factor has the limit
$R_{ij}\to(1+m_{j}^{2}/Q^{2})$, which we recognize as the
original slow-rescaling correction:
$$\bar{\xi} = \xi (1+m_j^2/Q^2),\quad\quad m_i=0.$$
Our focus in including quark mass corrections is for charm
production in neutrino scattering, where incident quark masses
are negligible, or where heavy quark PDFs are very small and
additionally suppressed by small mixing angles.
Consequently we shall focus on the $m_i=0$ results below.

%%%%%%%%%%%%%%%%%%%%%%%%%%%%%%%%%%%%%%%%%%%%%%%%%
\section{\patch Charm Mass Dependence in $h_2$ and $g_2$\label{app:charm}}

We discuss here the charm mass, $m_c$, dependence in the target mass
corrected structure functions in Eqs.~(\ref{eq:f1lo})--(\ref{eq:f3lo}).
Inserting $F_2^{(0)}$ from Eq.~(\ref{eq:f2c}) into the definition of
$h_{2}(\xi)$ in Eq.~(\ref{eq:defh2}), we find (for a diagonal CKM matrix):
\begin{eqnarray}
h_{2}(\xi)
&\equiv&
\int_{\xi}^{1}\der u\ \frac{F_{2}^{(0)}(u)}{u^{2}}
=\int_{\xi}^{1}\der u\ \frac{2\bar{u}s(\bar{u})}{u^{2}}
=2(1+m_{c}^{2}/Q^{2})\int_{\xi}^{1}\der u\ \frac{s(\bar{u})}{u}\ ,
\nonumber\\
\end{eqnarray}
where the variable $\bar{u}=u(1+m_{c}^{2}/Q^{2})$.
Making a variable transformation to $\bar{u}$, we find:
\begin{eqnarray}
h_{2}(\xi)
&=&
2(1+m_{c}^{2}/Q^{2})\int_{\xi}^{1}\der u\ \frac{s(\bar{u})}{u}
=2(1+m_{c}^{2}/Q^{2})
  \int_{\bar{\xi}}^{1+m_{c}^{2}/Q^{2}}\der\bar{u}\ \frac{s(\bar{u})}{\bar{u}}\ .
\nonumber\\
\end{eqnarray}
Since the strange quark PDF vanishes for $\bar{u}>1$ we can restrict
the upper integration bound to 1 and arrive at the final result:
\begin{equation}
\label{eq:h2lo}
h_{2}(\xi)=2(1+m_{c}^{2}/Q^{2})\int_{\bar{\xi}}^{1}\der u\ \frac{s(u)}{u}\ .
\end{equation}
Similar steps lead to the results:
\begin{eqnarray}
g_{2}(\xi) & = & 2\int_{\bar{\xi}}^{1}\der u\ \int_{u}^{1}\der v\ \frac{s(v)}{v}\ ,\label{eq:g2lo}\\
h_{3}(\xi) & = & \int_{\bar{\xi}}^{1}\der u\ \frac{2s(u)}{u}\ .\label{eq:h3lo}
\end{eqnarray}

%%%%%%%%%%%%%%%%%%%%%%%%%%%%%%%%%%%%%%%%%%%%%%%%%
\section{\patch Comparisons with the Literature\label{app:compare}}

In this appendix, we make the identification of our notation with the
notation in earlier papers by Georgi \& Politzer \cite{Georgi:1976ve}, 
De R\'ujula, Georgi \& Politzer \cite{DeRujula:1976tz}, and
Barbieri {\it et al.} \cite{Barbieri:1976rd}. We also comment on the
relation to the notation of the paper by Kretzer \& Reno \cite{Kretzer:2003iu}.
Known typographic errors are also indicated in this appendix, in an
attempt to minimize confusion when comparing results in the literature.
Since neutrino production of charm quarks has both target mass and
charm quark corrections, we make the comparison for charm production
in the diagonal Cabibbo matrix limit.

%%%%%%%%%%%%%%%%%%%%%%%%%%%%%%%%%%%%%%%%%%%%%%%%%
\subsection{\patch Comparisons with Georgi \& Politzer}

The results for the target mass corrected structure functions in
the work of  Georgi \& Politzer \cite{Georgi:1976ve} 
are given in their Eqs.~(4.19), (4.20) and (4.22).
Only leading order in QCD is considered, which in the massless limit
yields the Callan--Gross relation for the $F_1$ and $F_2$ structure
functions.  As a result, a function $F(\xi)$ is the only function introduced. 
Before proceeding, we should correct two typographic errors in
Ref.~\cite{Georgi:1976ve}: 
\begin{enumerate}
\item In relation (4.19) of Ref.~\cite{Georgi:1976ve}: 
\begin{eqnarray}
W_{1}(Q^{2},x)
&=&
\frac{x}{2(1+4M^{2}/Q^{2})^{1/2}}F(\xi)+...\to
\frac{x}{2(1+4x^{2}M^{2}/Q^{2})^{1/2}}F(\xi)
\nonumber\\
\end{eqnarray}
\item  In relation (4.22) of Ref.~\cite{Georgi:1976ve}: No derivation 
of $W_3$ is provided to get the overall factor, but there is
a factor 2 mismatch in relative normalization between the first and
second terms: 
\begin{eqnarray}
\frac{\nu W_{3}(Q^{2},x)}{M}
&=&
\frac{x}{2(1+4x^{2}M^{2}/Q^{2})}F(\xi)+...\to\frac{x}{(1+4x^{2}
M^{2}/Q^{2})}F(\xi)
\nonumber\\
\end{eqnarray}
\end{enumerate}

With these corrections, Eqs.~(4.19) and (4.20) of Ref.~\cite{Georgi:1976ve} 
agree with Eqs.~(\ref{eq:f1}) and (\ref{eq:f2}) at leading order if $F(\xi)$ in
\cite{Georgi:1976ve} is identified with the structure functions $F_{i}^{(0)}(\xi)$
such that:
\begin{equation}
F(\xi)=F_{2}^{(0)}(\xi)/\xi^{2}=2 F_1^{(0)}(\xi)/\xi\ .
\end{equation}
Note that $F_{2}^{(0)}(\xi)/\xi^{2}=2F_{1}^{(0)}(\xi)/\xi$ is the
Callan--Gross relation which holds in leading order for massless quarks.

The target mass corrected structure function $\nu W_{3}/M$ 
found in Eq.\ (4.22) in Ref.~\cite{Georgi:1976ve}, after
correcting the typographic error, agrees with Eq.~(\ref{eq:f3}) if:
\begin{equation}
F_3 = \frac{\nu}{2M} W_3\quad
{\rm and}\quad F(\xi)= 2 F_{3}^{(0)}(\xi)/\xi\ .
\end{equation}

In Ref.~\cite{DeRujula:1976tz} De R\'ujula {\em et al.} discuss the
electromagnetic structure functions, where $F(\xi)\rightarrow F_1^S(\xi)$
or $F_2^S(\xi)$. The $F_i^S$ are not defined, however, in terms of the
structure functions $F_i^{(0)}$. In addition, the structure functions $W_1$
and $W_2$ are a factor of $1/M$ times the $W_i$ defined in
Ref.~\cite{Georgi:1976ve}.

%%%%%%%%%%%%%%%%%%%%%%%%%%%%%%%%%%%%%%%%%%%%%%%%%
\subsection{\patch Comparisons with Barbieri, Ellis, Gaillard, \& Ross}

In comparing with the notation of Barbieri, Ellis, Gaillard, \& Ross \cite{Barbieri:1976rd},
we note firstly that the quantity ``$\nu$'' in \cite{Barbieri:1976rd} is scaled by a
factor $M$ compared to our notation, and we label it with a subscript $\nu_B$:
\begin{equation}
\nu_{B}=\nu M\ ,
\end{equation}
which gives $x=Q^2/(2\nu_B)=Q^2/(2M\nu)$.
Note also that the variable denoted by ``$\xi$'' in Ref.~\cite{Barbieri:1976rd}
corresponds to the generalized Nachtmann variable $\bar{\xi}$ used in this paper.

Substituting our notation in Eqs.~(2.12a--e) of Ref.~\cite{Barbieri:1976rd}
yields:
\begin{eqnarray}
W_{1}^B&=&(a^{2}+b^{2})\frac{x}{r}
\left[\left(1+\frac{m_{f}^{2}+m_{i}^{2}}{Q^{2}}\right)\mathcal{F}_{i}^B
(\bar{\xi})
+\mathcal{G}_{i}^B\right]\nonumber
\\ && -(a^{2}-b^{2})\frac{2 x m_{i}m_{f}}{Q^2 r}
\mathcal{F}_{i}^B(\bar{\xi})\ , 
\label{eq:barbW1}
\\
W_{2}^B&=& (a^{2}+b^{2})\frac{4 x^3 M^2}{r^3 Q^2}
\left[\left(\frac{R}{Q^{2}}\right)^{2}
\mathcal{F}_{i}^B(\bar{\xi})+3\mathcal{G}_{i}^B\right]\ , \\
W_{3}^B&=&2ab\frac{4 x^2 M^{2}}{r^2Q^2}
\left[\left(\frac{R}{Q^{2}}\right)^{2}\mathcal{F}_{i}^B(\bar{\xi})
+\frac{2 x M^{2}}{r Q^2}
\int_{\bar{\xi}}^{1}du\left(1-\frac{m_{i}^{2}}{M^{2}u^{2}}\right)
\mathcal{F}_{i}^B(u)\right]\ .
\label{eq:barbW3}
\nonumber\\
\end{eqnarray}
Here $R^2=[(Q^2+m_f^2-m_i^2)^2+4 m_i^2 Q^2]$,
$m_i$ and $m_f$ are the initial and final quark masses,
and $a$ and $b$ are the vector and axial-vector couplings,
respectively. 
The structure function  ${\cal F}_i$ corresponds to 
our $F_i^{(0)}$, and the 
 ${\cal G}_i$ term contains the single and double integral 
terms corresponding to our $h_i$ and $g_2$ expressions of 
Eq.~(\ref{eq:master}).

With $m_i^2\ll m_f^2$, in the first two structure functions
\begin{equation}
{(a^2+b^2)\mathcal{F}_i^B}(\bar{\xi}) = \frac{s(\bar{\xi})}{\bar{\xi}}
\end{equation}
yields $W_1^B$ and $W_2^B$ with the same normalization as
in Eq.~(\ref{eq:WtoF}). In the third structure function, we find that:
\begin{equation}
2 a b \mathcal{F}_i^B (\bar{\xi}) = \frac{s(\bar{\xi})}{\bar{\xi}}\ .
\end{equation}
We obtain agreement for the third structure function if
$\nu_B W_3^B = 2 M^2 F_3^{\TMC}$.

%%%%%%%%%%%%%%%%%%%%%%%%%%%%%%%%%%%%%%%%%%%%%%%%%

We note, however, that we do not match the expressions for Barbieri
{\it et al.} \cite{Barbieri:1976rd} in the limit of a 
finite initial quark mass ($m_i\not\simeq 0$).
While the initial quark mass terms are small numerically, for
completeness we comment on the relation of our results to this work.

The renormalization and factorization prescriptions allow
certain freedoms in organizing the terms in the master equation;
hence, the mass dependence need not be unique.
Our
results are constructed to a) reduce to standard parton model
expressions in various limiting cases ({\it e.g.} $\{m_i, m_f, M \}\to
0$), and to  b) maintain the simple structure of
Eq.~(\ref{eq:master}).
Our formulation incorporating quark masses follows the factorization
construction of Collins~\cite{Collins:1998rz}.
  In particular, the $\{A, B, C \}$ coefficients for our master formula, 
Eq.~(\ref{eq:master}), are free of any quark mass terms; the $m_i$
dependence is entirely contained in $F_i^{(0)}$.

Our expressions also satisfy the expected limiting cases.
For example, in the limit $M\to0$ our expressions reduce
to the general parton model result with finite $\{m_i,m_f\}$ terms, by
construction; this is in contrast to  Eqs.~(\ref{eq:barbW1})--(\ref{eq:barbW3}).

In summary, our expressions of Eq.~(\ref{eq:master}) follow the simple
factorized structure of Ref.~\cite{Collins:1998rz}, and reduce to the
standard parton model results in the limits of various vanishing
masses. The numerical difference between this formulation and that of
Ref.~\cite{Barbieri:1976rd}  is small.

%%%%%%%%%%%%%%%%%%%%%%%%%%%%%%%%%%%%%%%%%%%%%%%%%
%%%%%%%%%%%%%%%%%%%%%%%%%%%%%%%%%%%%%%%%%%%%%%%%%
\subsection{\patch Comparisons with Kretzer \& Reno}

In Kretzer and Reno \cite{Kretzer:2003iu}, the notation has
the slow rescaling variable including the charm quark mass
written as $\xi/\lambda$, where $\lambda = (1+m_c^2/Q^2)^{-1}$,
so that $\xi/\lambda=\bar{\xi}$.
The quantity $\rho$ in \cite{Kretzer:2003iu} equals $r$,
and $\mu=M^2/Q^2$.  In the absence of charm mass corrections,
the results of Ref.~\cite{Kretzer:2003iu} are in full agreement with
the equations in this paper.

The translation of the treatment of $F_i^{\TMC}$ in terms of
$F_i^{(0)}$ in Ref.~\cite{Kretzer:2003iu} is not made explicitly
when charm production is involved. Charm production is discussed
in terms of $\mathcal{F}_i\sim s$.  The expression for charm
production in Eq.~(3.27) of Ref.~\cite{Kretzer:2003iu} is written as
the sum of three terms times coefficients.  The terms depend on
$\mathcal{F}_i$, $\mathcal{H}_i$ and $\mathcal{G}_2$.
The quantity $\mathcal{F}_i(\xi)$ appearing in Eq.~(3.27)
should read $\mathcal{F}_i(\bar{\xi})$, with  
\begin{equation}
\mathcal{F}_i(\bar{\xi})=s(\bar{\xi})
\end{equation}
at leading order.
The functions $\mathcal{H}_i$ and $\mathcal{G}_2$ in
Ref.~\cite{Kretzer:2003iu} for charm production should also
be functions of $\bar{\xi}$ rather than $\xi$.
In terms of $g_2$ and $h_i$ introduced here, one has:
\begin{eqnarray}
h_2(\xi) &=& 2(1+m_c^2/Q^2)\mathcal{H}_2(\bar{\xi})\ , \\
g_2(\xi) &=& 2 \mathcal{G}_2(\bar{\xi})\ , \\
h_3(\xi) &=& 2\mathcal{H}_3(\bar{\xi})\ .
\end{eqnarray}
With these corrections, Eq.~(3.27) in Ref.~\cite{Kretzer:2003iu} is:
\begin{equation}
\label{eq:FKR}
F_j^{\TMC}(x,Q^2) \rightarrow \sum \alpha_j^i\mathcal{F}_i(\bar{\xi},Q^2)
+\beta_j^i\mathcal{H}_i(\bar{\xi},Q^2)
+\gamma_j \mathcal{G}_2(\bar{\xi},Q^2)\ .
\end{equation}
The sum is from $i=1, \cdots, 5$, however, there is no mixing of
the 4th and 5th terms in $F_{1,2,3}^{\TMC}$.  Tables~IV and V
with the coefficients $\alpha_j^i$ and $\beta_j^i$ are correct,
however, there is an error in Table~VI.  The coefficients $\gamma_i$
all have an extra power of $\lambda$ in the denominator.  Explicitly,
the first two coefficients should read (with the translated notation):
\begin{eqnarray}
\gamma_1 &=& \frac{4 M^4 x^3}{Q^4 r^3}\ , \\
\gamma_2 &=& \frac{24 M^4 x^4}{Q^4 r^5}\ .
\end{eqnarray}
In fact, Eq.~(\ref{eq:FKR}) was used in the numerical work in
Ref.~\cite{Kretzer:2003iu}. The contributions of $\gamma_1$ and
$\gamma_2$ are small, so the error in Table~VI is not numerically
significant.

%%%%%%%%%%%%%%%%%%%%%%%%%%%%%%%%%%%%%%%%%%%%%%%%%%%%%%%%%%%%%%%%%%%
%%%%%%%%%%%%%%%%%%%%%%%%%%%%%%%%%%%%%%%%%%%%%%%%%%%%%%%%%%%%%%%%%%%
%%%%%%%%%%%%%%%%%%%%%%%%%%%%%%%%%%%%%%%%%%%%%%%%%%%%%%%%%%%%%%%%%%%
\newpage{}
\section*{References}
\bibliographystyle{hunsrt}
\bibliography{tmc3}
%\bibliography{tmc}

\end{document}